\documentclass[journal,twoside,web]{ieeecolor}
\usepackage{generic}
\usepackage{graphics}
\usepackage{subfigure}
\usepackage{amsfonts}
\usepackage{mathtools}
\usepackage{mathrsfs}
\usepackage{epsfig}
\usepackage{times}
\usepackage{amsmath}
\usepackage{amssymb}
\usepackage{algorithm}
\usepackage{ulem}
\usepackage{mathabx}
\usepackage{enumerate}
\usepackage{hyperref}
\usepackage{wrapfig}
\newtheorem{theorem}{Theorem}
\newtheorem{corollary}{Corollary}
\newtheorem{definition}{Definition}

\newtheorem{lemma}{Lemma}
\newtheorem{remark}{Remark}
\newtheorem{assumption}{Assumption}
\def\BibTeX{{\rm B\kern-.05em{\sc i\kern-.025em b}\kern-.08em
 T\kern-.1667em\lower.7ex\hbox{E}\kern-.125emX}}
\begin{document}
\title{\bf Differentially Private Recursive Least Squares Estimation for ARX Systems
with Multi-Participants \thanks{The work is supported by National Natural Science Foundation of China under Grant No. 62203045, 62433020 and T2293770. Corresponding author: Jimin Wang.}}
\author{Jianwei Tan, Jimin Wang, \IEEEmembership{Member,~IEEE,} and Ji-Feng Zhang, \IEEEmembership{Fellow,~IEEE}
\thanks{Jianwei Tan is with the Key Laboratory of Systems and Control, Institute of Systems Science, Academy of Mathematics and Systems Science, Chinese Academy of Sciences, Beijing 100190, and also with the School of Mathematical Sciences, University of Chinese Academy of Sciences, Beijing 100049, China. (e-mail: jwtan@amss.ac.cn)}
\thanks{Jimin Wang is with the School of Automation and Electrical Engineering, University of Science and Technology Beijing, Beijing, 100083, China and also with the Key Laboratory of Knowledge Automation for Industrial Processes, Ministry of Education, Beijing 100083, China (e-mail: jimwang@ustb.edu.cn)}
\thanks{Ji-Feng Zhang is with the School of Automation and Electrical Engineering, Zhongyuan University of Technology, Zheng Zhou 450007; and also with the Key Laboratory of Systems and Control, Institute of Systems Science, Academy of Mathematics and Systems Science, Chinese Academy of Sciences, Beijing 100190, China. (e-mail: jif@iss.ac.cn)}}

\maketitle
\begin{abstract}
This paper proposes a differentially private recursive least squares algorithm to estimate the parameter of autoregressive systems with exogenous inputs and multi-participants (MP-ARX systems) and protect each participant's sensitive information from potential attackers. We first give a rigorous differential privacy analysis of the algorithm, and establish the quantitative relationship between the added noises and the privacy-preserving level when the system is asymptotically stable. The asymptotic stability of the system is  necessary for ensuring  the differential privacy of the algorithm. We then give an estimation error analysis of the algorithm under the general and possible weakest excitation condition without requiring the boundedness, independence and stationarity on the regression vectors. Particularly, when there is no regression term in the system output and the differential privacy only on the system output is considered, $\varepsilon$-differential privacy and almost sure convergence of the algorithm can be established simultaneously. To minimize the estimation error of the algorithm with $\varepsilon$-differential privacy,  the existence of  the noise intensity is proved. Finally, two examples are given to show the efficiency of the algorithm.
\end{abstract}

\IEEEpeerreviewmaketitle
\section{Introduction}
As one of the most fundamental methods for data analysis and system identification, the least-squares method  has made a series of theoretical achievements \cite{moore1978strong}-\cite{Sayedana2024}, and has been ubiquitously employed in numerous fields, such as engineering systems, physical systems, social systems, biological systems, economic systems and many others \cite{chen2014recursive}-\cite{ljung1987}. For example, the least-squares method is used to study the relationship between the environment in which children grow up and their prospect for economic mobility \cite{chetty2018opportunity}, to find that how solar radiation
and pump velocity affect the temperature in heat storage, and to construct a mathematical model for aircraft's dynamic behaviors \cite{ljung1998system}. In practice, the system inputs may be provided by more than one participant, such as in game theory  \cite{xu2021stackelberg}, engineering \cite{perreault2005optimal} and federated learning \cite{yang2019federated}. Besides, the existence of the regression terms also makes the model suitable for describing control systems where the historical output usually influences present output. One of the examples is the disease-related diagnosis in hospitals. A patient may go to different hospitals for different diseases in seeking of better treatment. These hospitals want to study the relationship between these different diseases, and a natural way is to share patients' information. Here the disease of interest can be viewed as the system output, and other related disease can be viewed as the system input that provided by different hospitals. In this case, the MP-ARX systems can naturally be used to model the relationship of those diseases. Hospitals need to protect the privacy of individual patients. However, the privacy of each participant may be breached when each participant sends its data directly to date center if the potential attackers exist. So the data involved in the least squares method may be sensitive and need to be protected. In addition to the patient medical records just mentioned, these sensitive information may also be personal income in social research \cite{chetty2018opportunity}, medical records in drug administration \cite{aastrom2008feedback}, or power consumption in householder \cite{lisovich2010inferring}. If the sensitive information is leaked, then it may threaten property, privacy and even life. Therefore, it is of great importance to develop a privacy-preserving estimation method that can not only realize the benefits of the least-squares method but also protect the sensitive information involved.

Recently, a brief but comprehensive discussion for privacy security in control systems has been given in \cite{zhang2021privacy}. Up to now, there are many techniques developed to protect the privacy in control systems, such as structure based techniques \cite{wang2019privacy}, isomorphic transformation \cite{sultangazin2021symmetries}, homomorphic encryption \cite{xu2018information,tjell2019privacy,Lu2018data,Ruan2019} and stochastic perturbation \cite{LeNy2014differentially}-\cite{Nekouei2022}. Structure based techniques and isomorphic
transformation method usually require the data center directly access the sensitive information. These kind of methods are not applicable to privacy protection of the least-squares method when the data provider do not trust the data center. Homomorphic encryption method allows addition or multiplication similar to plaintext in ciphertext, but this kind of method
requires a large amount of computation and is also time-consuming, which limits its use in practice. In addition, homomorphic encryption based privacy preserving recursive least-squares methods usually require that at least one server is trustworthy \cite{tjell2019privacy}.

In the stochastic perturbation technique, differential privacy and correlated noises are two commonly used methods.  Assuming that potential attackers cannot access the entire neighborhood set of an agent, a well-designed correlated noise sequence has been used in \cite{Mo2017} to obfuscate the sensitive information. As pointed out in \cite{Ruan2019}, if potential attackers obtain the information received and delivered by an agent, then this agent’s initial state can be estimated through an iterative observer under such correlated noises mechanism. Different from the correlated noises privacy-preserving approach, differential privacy possesses a more robust privacy of an agent regardless of any auxiliary information potential attackers may have. Besides, differential privacy is easy to implement and computationally lightweight,  and thus becomes a de facto standard for privacy protection \cite{dwork2006calibrating,dwork2006differential,dwork2014algorithmic,Liu2018}. Up to now, the least-squares method with differential privacy requirements has been
studied from different aspects, such as differentially private M-estimate \cite{lei2011differentially}, differentially private empirical risk minimization
\cite{chaudhuri2011differentially, kifer2012private, bassily2014private}, differentially private linear regression \cite{zhang2012functional}-\cite{Liu2024}. Differentially private M-estimate \cite{lei2011differentially} and empirical risk minimization \cite{chaudhuri2011differentially, kifer2012private, bassily2014private} were presented, but the regression vector was required to be bounded. A differentially private iteratively reweighted least-squares algorithm was developed without considering the convergence analysis
\cite{park2016note}. A differentially private algorithm for linear regression learning in a decentralized fashion was presented, but the estimation error increases linearly or exponentially \cite{liuyang2020differentially}. The differential privacy for linear regression by the functional mechanism \cite{zhang2012functional}, high dimensional sparse linear regression \cite{wang2018high, wang2019sparse} and linear regression with unbounded covariance \cite{milionis2022differentially} were studied, respectively. However, the regressor vector therein has independent identically distributed requirements, which is not suitable for the systems with regression terms.

In this paper, we focus on a differentially private recursive least-squares algorithm for MP-ARX systems. The $\varepsilon$-differential privacy analysis of the algorithm is given for each participant. Then, the estimation error of the algorithm are provided under the general and possible weakest excitation condition. The main contributions of this paper are as follows:
\begin{itemize}
\item A differentially private recursive least-squares algorithm for MP-ARX systems is proposed to estimate unknown parameters and protect each participant's sensitive information from potential attackers. Unlike most existing works, the system considered is with inputs  distributed in different participants, and is more general than the ARX model~ \cite{chen2014recursive, ljung1998system}.
\item The quantitative relationship between the added noises and the privacy level $\varepsilon$ is obtained for each participant. Besides, it is proved that the asymptotic stability of the system is necessary for ensuring the differential privacy of the algorithm.
\item The estimation error of the algorithm is given under the general and possible weakest excitation condition without requiring the boundedness, independence and stationarity on the regression vectors. More interestingly, when there is no regression term in the system output and the differential privacy only on the system output is considered, the  almost sure convergence and $\varepsilon$-differential privacy of the algorithm is obtained simultaneously.
\item When the algorithm is $\varepsilon$-differentially private,   the existence  of the noise parameters is proved to minimize the estimation error of the algorithm.
\end{itemize}
The rest of this paper is organized as follows. Problem formulation is given in Section 2. The proposed algorithm, the privacy and estimation error analysis of the algorithm are presented in Section 3. Two examples are given to show the efficiency of the algorithm in Section 4, and some concluding remarks are given in Section 5.

{\it Notation:} In this paper, $\mathbb{C}$, $\mathbb{R}$, $\mathbb{N}$,
$\mathbb{R}^{n}$ denote the sets of complex numbers, real numbers,
non-negative integers and $n$-dimension Euclidean space, respectively. $\mathbb{R}^{m\times n}$ and $\mathscr{B}(\mathbb{R}^n)$ denote the set of $m$-by-$n$ real matrix and Borel sets on $\mathbb{R}^n$, respectively. $(\mathbb{R}^+)^n =\{(x_1,\ldots,x_n)|x_i>0, 1\leq i\leq n\}$. For $x \in \mathbb{R}^n$, define $\|x \|_1 = \sum_{i=1}^n |x_i|,
\|x\| = \sqrt{\sum_{i=1}^n x_i^2}$. For $M \in \mathbb{R}^{m\times n}$,
define $\|M\| =\sup_{\|x\|=1}\|Mx\|$ with $x\in \mathbb{R}^n$. Further,
if matrix $M$ is symmetric and semi-positive definite, define
$\lambda_{\min}(M)$ and $\lambda_{\max}(M)$ as the smallest and largest
eigenvalue of $M$, respectively. For sequences $\{x_k\}$ and $\{x_{i,k}\}$ with $x_k, x_{i,k}\in \mathbb{R}$, and any two positive integers $t\geq s$, denote $x_{[s:t]}=[x_{s}, x_{s+1},\ldots, x_{t}]^T,$ $x_{i,[s:t]}=[x_{i,s}, x_{i,s+1}, \ldots, x_{i,t}]^T$. For $D \in \mathscr{B}(\mathbb{R}^{T})$ with $(x_1, x_2, \dots, x_{T}) \in D$, denote $D|_{x_k} = \{x_k|(x_1, \dots, x_k, \dots, x_T)\in D \}$. Given two real valued functions $f(n)$ and
$g(n)$ defined on $\mathbb{N}$ with $g(n)$ being strictly positive for
sufficiently large $n$, denote $f(n)=O(g(n))$ if there exist $ M > 0$
and $n_0 > 0$ such that $|f(n)|\leq Mg(n)$ for
any $ n\geq n_0$; $f(n) = o(g(n))$ if for any $ \varepsilon >0$ there
exists $n_0$ such that $|f(n)|\leq \varepsilon g(n)$ for any $n > n_0$.
$L(\mu, b)$ denotes the Laplacian distribution with probability density
distribution $f(x|\mu, b) = \frac{1}{2b}\exp(-\frac{|x-\mu|}{b})$, mean $\mu$ and variance $2b^2$.
\section{Problem formulation}
Consider the following MP-ARX systems:
\begin{align}
\label{eq:DP_LS_system}
y_{k+1}&=a_1y_k+a_2y_{k-1}+\cdots+a_py_{k+1-p} \notag	\\
& +b_{1,1}u_{1,k}+b_{1,2}u_{1,k-1}+\cdots+b_{1,q_1}u_{1,k+1-q_1}
\notag \\
&  +b_{2,1}u_{2,k}+b_{2,2}u_{2,k-1}+\cdots+b_{2,q_2}u_{2,k+1-q_2}
\notag \\
&  \vdots \notag\\
& +b_{m,1}u_{m,k}+b_{m,2}u_{m,k-1}+\cdots+b_{m,q_m}u_{m,k+1-q_m}
\notag \\
& +\omega_{k+1}, \quad \forall k\in \mathbb{N},
\end{align}
where $p \in \mathbb{N}, q_{i} \in \mathbb{N}$ for $i=1,\dots,m$ are known
system orders; $a_j \in \mathbb{R}, b_{i,l}\in \mathbb{R}$,
for $l=1,\dots,q_i, j=1,\dots,p, i=1,\dots,m$ are unknown parameters;
$y_k\in \mathbb{R}$ is the system output measured by Participant
$\mathcal{P}_{0}$ at time $k$, $u_{i,k}\in \mathbb{R}$ is the system input provided by Participant
$\mathcal{P}_i$ at time $k$ for $i=1,\dots,m$;
$\omega_k \in \mathbb{R}$ is the system noise at time $k$.
Without loss of generality, we assume $y_{k+1} = 0$ and $u_{i,k}=0$ for
$k\leq 0$.

For compact expression, we introduce the following vectors:
\begin{align*}
\theta &= [a_1, \ldots ,a_p, b_{1,1}, \ldots, b_{1,q_1},
\ldots,b_{m,1}, \ldots, b_{m,q_m}]^T,
\cr
\varphi_k &= [y_{k},\ldots, y_{k+1-p}, u_{1,k},
\ldots, u_{1,k+1-q_1}, \cr
&\quad \ldots,
u_{m,k}, \ldots, u_{m,k+1-q_m}]^T,
\end{align*}
and rewrite \eqref{eq:DP_LS_system} as:
\begin{align}
y_{k+1} = \theta^T\varphi_{k}+\omega_{k+1}, \quad \forall k\in \mathbb{N}.
\label{eq:DP_LS_system_compressed}
\end{align}

\begin{figure}[httb]
\centering
\includegraphics[width=0.4\textwidth]{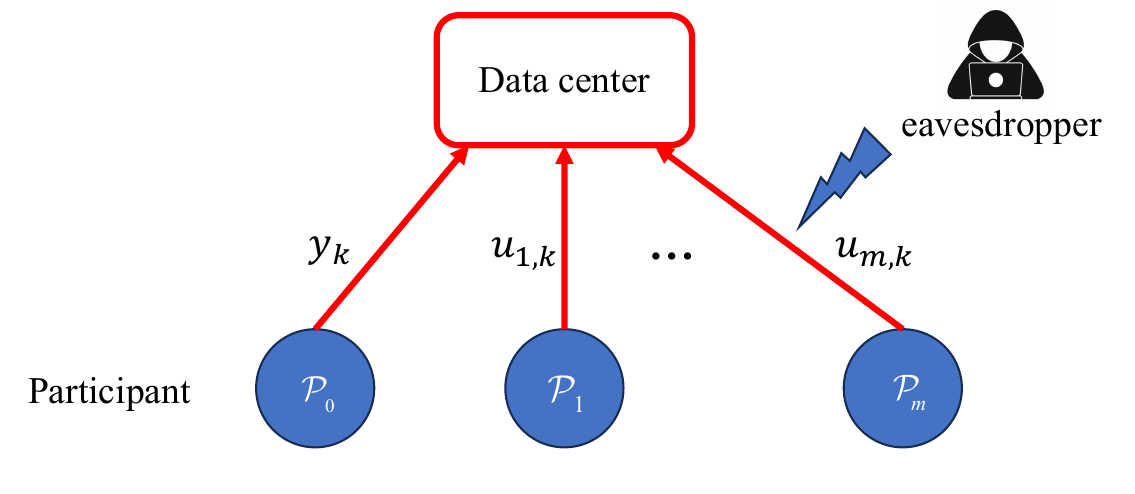}
\caption{Architecture of the problem: multiple participants (i.e. $\mathcal{P}_0$, $\cdots$, $\mathcal{P}_m$ ) collaborate to complete a parameter identification problem that cannot be completed by any individual participant, where participants are low-resource parties with sensitive information that they outsource to a powerful data center (may be one participant). The data center has to solve an identification problem on the sensitive information of the participants}
\end{figure}

The situation we consider in this paper is that a data center (may be one participant)
wants to collect information from the involved participants to identify the
unknown parameters in System \eqref{eq:DP_LS_system}, and each
participant protects its sensitive information against the potential attacker. The architecture of the problem is presented in Fig. 1. Participant $\mathcal{P}_0$'s sensitive information is $y_{[1:T_1]}$, and Participant $\mathcal{P}_i$'s ($i=1, \cdots, m)$ sensitive information is $u_{i,[0:T_1-1]}$ for any $T_1\geq1$. The potential attacker may be the semi-honest data center or an
eavesdropper \cite{zhang2021privacy} with the following abilities:
\begin{itemize}
\item Eavesdropping on all communication channels between each participant and the data center.
\item Colluding with some participants to infer other participants' sensitive information.
\end{itemize}

In this case, participants' sensitive information will be leaked to the potential attackers if they send information directly to the data center. To address this issue, we propose a differentially private recursive least-squares estimation algorithm.

Next, we will provide two econometric research examples on the scenarios considered in this paper.

The first one is on the impact study of historical investment behaviors of various banks in economic development situation. $y_{k}$ represents the economic development situation of the $k$th month (internal evaluation indicators), and $u_{i,k}$, $i=1,\cdots,m$ represents the credit investment scale of the $i$th bank in the $k$th month.  System \eqref{eq:DP_LS_system} is used to identify the relationship between the economic development situation of the $k+1$th month and historical economic development situation and the historical investment situation of each bank. Among them, data about the economic development situation and the credit investment scale of various banks need to be protected.

The second one is on the impact study of local debt resolution investment in various provinces on overall systemic risk governance.  $y_{k}$ represents the systemic risk indicator for the $k$th month, and $u_{i,k}$, $i=1,\cdots,m$ represents the government monetary policy investment received by the $i$th province in the $k$th month. System \eqref{eq:DP_LS_system} is used to identify the relationship between the systemic risk of the $k+1$th month and cumulative systemic risk level  and  recent government investments in each province. Among them, systematic risk indicators and local debt resolution allocation quota data in various provinces need to be protected.
\begin{remark}
When $m=1$, the architecture of the problem becomes  two participants (i.e. $\mathcal{P}_0$ and $\mathcal{P}_1$ ) collaborate to identify parameters while protecting each participant's sensitive information. In this case, System \eqref{eq:DP_LS_system} reduces to the normal ARX model \cite{chen2014recursive, ljung1998system}.  So the results of this paper  are also applicable to the privacy issues involved in the normal ARX model. 
\end{remark}
\begin{remark}
The architecture of the studied problem is similar to the federated learning, and adheres to (i) privacy of data, (ii) local computing, and, (iii) data transmission.  Federated learning is a learning architecture where multiple devices (participants) collaborate to train a model under the coordination of a data central. The learning architecture adheres to (i) privacy of data, (ii) local computing, and, (iii) model transmission \cite{yang2019federated}.  Thus, the  system inputs and outputs are exchanged in this paper,  while the models are exchanged in federated learning.  Besides, in this paper, local computing only occurs in the data center while it occurs in each participant including the data center in federated learning.
\end{remark}
\section{Main result}
\subsection{Algorithm}
For each participant, let
\begin{align}
\mathcal{P}_{0} &:\,\,\, \bar{y}_k =  y_k + \eta_k
\label{eq:DP_LS_ybar}, \\
\mathcal{P}_i &: \bar{u}_{i,k} =  u_{i,k} + \xi_{i,k},
\quad i=1, \cdots, m,
\label{eq:DP_LS_ubar}
\end{align}
where $ \eta_k \sim L(0,b_0),\, \xi_{i,k} \sim  L(0, b_i)$, $b_0$
and $b_i$ are the parameters to be designed. At time $k$, Participant $\mathcal{P}_{i}$ sends $\bar{u}_{i,k-1}$ to the data center, and Participant $\mathcal{P}_{0}$ sends
$\bar{y}_k$ to the data center.

For the data center, let
\begin{align*}
\bar{\varphi}_k &= [\bar{y}_k, \ldots, \bar{y}_{k-p+1},
\bar{u}_{1,k}, \ldots, \bar{u}_{1,k-q_1+1}, \cr
& \quad \ldots, \bar{u}_{m,k}, \ldots, \bar{u}_{m,k-q_m+1}]^\mathrm{T}
\in \mathbb{R}^{p+\sum_{i=1}^mq_i},
\end{align*}
and at time $k+1$,
\begin{align}
\bar{a}_k&=(1+\bar{\varphi}_k^T\bar{P}_k\bar{\varphi}_k)^{-1},
\label{eq:DP_LS_alg_theta}\\
\theta_{k+1} &= \theta_{k}+\bar{a}_k\bar{P}_k\bar{\varphi}_k(\bar{y}_{k+1}
- \bar{\varphi}_k^T\theta_{k}),
\label{eq:DP_LS_alg_a}\\
\bar{P}_{k+1}&=(\bar{P}_{0}^{-1}+\sum_{i=0}^k \bar{\varphi}_i^T
\bar{\varphi}_i)^{-1}
=\bar{P}_{k}-\bar{a}_k\bar{P}_k\bar{\varphi}_k\bar{\varphi}_k^T\bar{P}_k,
\label{eq:DP_LS_alg_P}
\end{align}
where $\theta_k$ is the estimate of $\theta$ at time $k$,
$\bar{P}_0^{-1} = \alpha I$ for any given small positive
$\alpha$, $\theta_0$ is the initial value of $\theta$
at time $k=0$.
\begin{remark}
The recursive algorithm \eqref{eq:DP_LS_alg_theta}-\eqref{eq:DP_LS_alg_P}  updates the estimate as soon as the new data are collected.  As shown in \cite{ljung1987}, the recursive algorithm is online  and  a key instrument in adaptive control, adaptive filtering, and adaptive  prediction. This can be applied to many practical problems, such as ship steering, short-term prediction of power demand, digital transmission of speech, channel equalization, monitoring and failure detection  \cite{ljung1987}. Furthermore, the proposed algorithm consumes less storage space and updates the estimate step by step as the data grows,  and hence protect a more robust privacy of each participant.
\end{remark}
\begin{remark}
Different from the recursive least-squares algorithm {\rm \cite{chen2014recursive}}, the perturbed information $\bar{y}_k$ and $\bar{u}_{i,k}$ are used to protect the privacy. It is worth noting that
noises are added before information is sent out. Thus, the attacker's available information to infer the sensitive information can only be $\bar{y}_k$, $\bar{u}_{i,k}$ and the raw information of those participants that the attacker colludes with.
\end{remark}
\begin{remark}
The estimation of multi-participant systems is more difficult compared with that of multi-agent systems (e.g. \cite{Liu2018}), because each participant cannot identify the unknown parameters by only using its own data. This makes the communications more important for the multi-participant systems, and further makes the data more susceptible to leakage. Therefore,  privacy protection in multi-participant systems is more necessary than that in multi-agent systems.
\end{remark}

\subsection{Privacy analysis}
In this subsection, we first give sufficient conditions to establish $\varepsilon$-differential privacy of each participant, which reveals the relationship of the privacy level and the added noise. To do so, inspired by \cite{LeNy2014differentially,Nozari2017}, we first introduce the following concepts on differential privacy.

\begin{definition}[$\delta$-adjacency]
Given a positive real number $\delta > 0$ and a distance space
$(X, d)$. For any $x, y \in X$, we say that $x$ and $y$ are
$(\delta, d)$-adjacent or $\delta$-adjacent under distance
$d(\cdot,\cdot)$  if $d(x,y)\leq \delta.$ Specifically, if $X=\mathbb{R}^n, d(x,y)=\|x-y\|_1= \sum_{i=1}^n |x_i-y_i|\leq \delta$, then we say that $x$ and $y$ are $\delta$-adjacent.
\end{definition}

\begin{definition}[$\varepsilon$-differential privacy] \label{def:DP_LS_DP}
Let $\varepsilon, \delta$ be two positive numbers,
$(X,d)$  be a distance space of sensitive information,
$(Y, \mathscr{B}(Y))$ be a measurable space of the attacker's observed information,
$(\Omega, \mathscr{F}, \mathbb{P})$ be  a probability space,
$\mathcal{M}: X \times \Omega \to Y$ be a random mechanism.
If for all $\delta$-adjacent $x_1, x_2 \in X$ and for all $D\in \mathscr{B}(Y)$,
\begin{align}
\mathbb{P}({\mathcal{M}(x_1)\in D}) \leq e^{\varepsilon}
\mathbb{P}({\mathcal{M}(x_2)\in D}),
\label{eq:DP_LS epsion}
\end{align}
then we say that $\mathcal{M}$ is $\varepsilon$-differentially private under
$\delta$-adjacency.
\end{definition}

\begin{remark}
 (\ref{eq:DP_LS epsion}) is standard in defining the differential privacy. Since it holds for all $\delta$-adjacent $x_1, x_2 \in X$, we can exchange $x_1$ with $x_2$ and obtain $\mathbb{P}({\mathcal{M}(x_2)\in D}) \leq e^{\varepsilon} \mathbb{P}({\mathcal{M}(x_1)\in D})$. Since $e^{\varepsilon}\approx 1+\varepsilon$ for a small $\varepsilon > 0$, it means that for a sufficiently small $\varepsilon > 0$, the attacker cannot distinguish $x_1$  from $x_2$  based on the observation $D$, and hence we say that the $\varepsilon$-differential privacy is achieved.
\end{remark}

\begin{remark}
Both $\varepsilon$-differential privacy and $(\varepsilon, \nu)$-differential privacy can be used to protect the sensitive information of each  participant. Specifically, the Laplacian noise is used in $\varepsilon$-differential privacy, which is analyzed by an $L_{1}$ norm, while the Gaussian noise is used in $(\varepsilon, \nu)$-differential privacy, which is analyzed by an $L_{2}$ norm.  For the definition of $(\varepsilon, \nu)$-differential privacy, please refer to \cite{LeNy2014differentially,dwork2014algorithmic}.  In order to clearly express our goal and calculate conveniently, we choose $\varepsilon$-differential privacy as the privacy-preserving method in our paper. Note that if $(\varepsilon, \nu)$-differential privacy is used, then the privacy and convergence analysis of the algorithm still holds. 
\end{remark}

\begin{remark}
Intuitively, $\delta$ characterizes the ``closeness" of two pieces of similar
sensitive information, $\varepsilon$ characterizes the difficulty
for an attacker to distinguish the sensitive information $x_1$ from
its similar neighbor $x_2$ based on the observed information
(some $D\subset Y$). The smaller $\varepsilon$ is, the more difficult it is
for an attacker to distinguish. Hence, smaller $\varepsilon$ and larger
$\delta$ means a better privacy protection level.
\end{remark}

Next, we introduce the following assumptions and lemma for privacy analysis.
\begin{assumption}\label{ass:DP_LS_A1}
System \eqref{eq:DP_LS_system} is asymptotically stable, i.e. $\lambda(z) := 1 - a_1z - a_2 z^2 - \dots - a_pz^p \neq 0, \,
\forall \, |z| \leq 1$.
\end{assumption}
\begin{assumption}\label{ass:DP_LS_A2}
$\{\xi_{i,k}\}, i=1,\cdots, m, \{\eta_{k}\}$ are mutually independent of each other, and each sequence consists of independent elements. Besides, $\xi_{i,k}$ and $\eta_{k+1}$ are all independent of $\{y_k\}, \{u_{i,k}\}$ and $\{\omega_k\}$.
\end{assumption}

\begin{lemma}\label{remark lS 2}
If Assumption~\ref{ass:DP_LS_A1} holds,  then there exist constants $c_0>0$
and $\lambda\in(0,1)$ such that
\begin{align}\label{eq:DP_LS_c0}
\| A^k \| \leq c_0\lambda^k, \quad \forall k\in\mathbb{N},
\end{align}
where 
\begin{align}
A & = \begin{bmatrix}
    0       & 1     &  0        & \cdots & 0        & 0     \\
    0       & 0     &  1        & \cdots & 0        & 0     \\
    \vdots  & \vdots&  \vdots   & \cdots & \vdots   & \vdots \\
    0       & 0     &  0        & \cdots & 0        & 1     \\
    a_p     &a_{p-1}&  a_{p-2}  & \cdots & a_2      & a_1
    \end{bmatrix}.
\label{eq:DP_LS_matrix_A}
\end{align}
\end{lemma} 
\noindent
{\bf Proof}.  By Gelfand formula in \cite{Horn2012} and Assumption~\ref{ass:DP_LS_A1}, the lemma is proved.   \hfill $\blacksquare$
\begin{theorem}(Differential privacy for Participant $\mathcal{P}_{0}$)
\label{thm:DP_LS_thm_1}
For System \eqref{eq:DP_LS_system} and Algorithm
\eqref{eq:DP_LS_alg_theta}-\eqref{eq:DP_LS_alg_P}, let
$\delta > 0$ and $\varepsilon > 0$ be any prescribed
privacy indexes. If Assumptions~\ref{ass:DP_LS_A1} and \ref{ass:DP_LS_A2}
hold, and Participant $\mathcal{P}_0$ chooses $b_0$ such that
\begin{align}
\frac{C_1}{b_0}\delta \leq \varepsilon,
\label{eq:DP_LS_thm_1_b0}
\end{align}
where
\begin{align}
C_1=1+\frac{\sqrt{p}c_0\lambda}{1-\lambda}
\label{eq:DP_LS_C1}
\end{align}
with $c_0$ and $\lambda$ satisfying \eqref{eq:DP_LS_c0},
then Algorithm \eqref{eq:DP_LS_alg_theta}-\eqref{eq:DP_LS_alg_P} is
$\varepsilon$-differentially private under $\delta$-adjacency for
Participant $\mathcal{P}_{0}$.
\end{theorem}
\noindent
{\bf Proof}. Note that the system may operate for a long time that exceeds $T_1$, and the existence of the regression terms makes $y_k$ with $k\geq T_1$ carry former outputs' information.
Then, the observed information for the attacker is $\bar{y}_{[1:T_2]}$, where $T_2 \geq T_1$ is the attack duration. We denote this correspondence as
$\mathcal{M}_1(y_{[1:T_1]}) = \bar{y}_{[1:T_2]}.$
It suffices to show that for any $T_2 \geq T_1$, any $y_{[1:T_1]}$ and
$y'_{[1:T_1]}$ satisfying $\|y_{[1:T_1]}-y'_{[1:T_1]}\|_1\leq \delta$,
and any $D\in \mathscr{B}(\mathbb{R}^{T_2})$, we have
$\mathbb{P}({\mathcal{M}_1(y_{[1:T_1]})\in D}) \leq e^{\varepsilon}
\mathbb{P}({\mathcal{M}_1(y'_{[1:T_1]})\in D}).$

Denote $u_{[0:T_{2}-1]}=[u^{T}_{1,[0:T_{2}-1]},\cdots,u^{T}_{m,[0:T_{2}-1]}]^T$.
Then, $y_{[T_1+1:T_2]}$ is determined by
$y_{[1:T_1]},u_{[0:T_2\!-\!1]},w_{[T_1\!+\!1:T_2]}$
according to System \eqref{eq:DP_LS_system}.
Note that $\mathbb{P}({\mathcal{M}_1(y_{[1:T_1]})\in D}) =
\mathbb{P}({\bar{y}_{[1:T_2]}\in D|y_{[1:T_1]}}).$ Then, by Assumption~\ref{ass:DP_LS_A2},
we have
\begin{align}
&\quad \mathbb{P}({\mathcal{M}_1(y_{[1:T_1]})\in D|
u_{[0:T_2-1]}, w_{[T_1+1:T_2]}})
\notag \\
&=\mathbb{P}({\bar{y}_{[1:T_2]}\in D|y_{[1:T_1]}, u_{[0:T_2-1]}, w_{[T_1+1:T_2]}})
\notag \\
&=\int_{\mathbb{R}^{T_2}}
(\frac{1}{2b_0})^{T_2} \cdot \notag
\prod_{k=1}^{T_2}1_{D|_{\bar{y}_k}}(y_k+\eta_k)
e^{-\frac{|\eta_k|}{b_0}} \mathrm{d}\Lambda_{T_2}
\notag \\
&= (\frac{1}{2b_0})^{T_2}
\int_{\mathbb{R}^{T_2}} \prod_{k=k_0}^{T_2}1_{D|_{\bar{y}_k}}(\bar{y}_k)
e^{-\frac{|\bar{y}_k-y_k|}{b_0}} \mathrm{d} \bar{\Lambda}_{T_2},
\label{eq:DP_LS_thm_1_proof_1}
\end{align}
where $\mathrm{d}\Lambda_{T_2}=\mathrm{d}\eta_1\cdots\mathrm{d}\eta_{T_2},$
$\mathrm{d}\bar{\Lambda}_{T_2}=\mathrm{d}\bar{y}_1\cdots\mathrm{d}\bar{y}_{T_2}.$

Similarly, one can get
\begin{align}
&\mathbb{P}({\mathcal{M}_1(y'_{[1:T_1]})\in D| u_{[0:T_2-1]}, w_{[T_1+1:T_2]}})
\notag \\
=& (\frac{1}{2b_0})^{T_2}
\int_{\mathbb{R}^{T_2}} \prod_{k=k_0}^{T_2}1_{D|_{\bar{y}_k}}(\bar{y}_k)
e^{-\frac{|\bar{y}_k-y'_k|}{b_0}} \mathrm{d} \bar{\Lambda}_{T_2},
\label{eq:DP_LS_thm_1_proof_2}
\end{align}
When $u_{[0:T_2-1]}$ and $w_{[T_1+1:T_2]}$ are given in System
\eqref{eq:DP_LS_system}, by Assumption~\ref{ass:DP_LS_A1},
we have
\begin{align}
&\sum_{k=1}^{T_2}|y_k-y'_k|
=
\sum_{k=1}^{T_1}|y_k-y'_k| + \sum_{k=T_1+1}^{T_2}|y_k-y'_k|
\notag \\
\leq&
\sum_{k=1}^{T_1}|y_k-y'_k| + \sum_{k=T_1+1}^{T_2}\|y_{[k+1-p:k]}-y'_{[k+1-p:k]}\|_1
\notag \\
=&
\delta + \sqrt{p}\sum_{k=T_1+1}^{T_2}\|A^{k-T_1}(y_{[T_1+1-p:T_1]}-y'_{[T_1+1-p:T_1]})\|
\notag \\
\leq&
(1+\sqrt{p}\frac{c_0\lambda}{1-\lambda})\delta
\triangleq C_1\delta,
\label{eq:DP_LS_thm_1_proof_3}
\end{align}
where $A$ is defined in \eqref{eq:DP_LS_matrix_A}.

Note that
\begin{align*}
&\quad\,
\int_{\mathbb{R}^{T_2}} \prod_{k=k_0}^{T_2}1_{D|_{\bar{y}_k}}(\bar{y}_k)
e^{-\frac{|\bar{y}_k-y_k|}{b_0}} \mathrm{d} \bar{\Lambda}_{T_2}
\notag \\
&\leq
\int_{\mathbb{R}^{T_2}} \prod_{k=1}^{T_2}1_{D|_{\bar{y}_k}}(\bar{y}_k)
e^{-\frac{|\bar{y}_k-y'_k|-|y_k-y'_k|}{b_0}} \mathrm{d}
\bar{\Lambda}_{T_2}
\notag \\
&=
e^{\frac{\sum_{k=1}^{T_2} |y_k-y'_k|}{b_0}}
\int_{\mathbb{R}^{T_2}}\prod_{k=1}^{T_2}1_{D|_{\bar{y}_k}}(\bar{y}_k)
e^{-\frac{|\bar{y}_k-y'_k|}{b_0}} \mathrm{d}\bar{\Lambda}_{T_2}.
\end{align*}
Then, by \eqref{eq:DP_LS_thm_1_b0}, \eqref{eq:DP_LS_thm_1_proof_1}-\eqref{eq:DP_LS_thm_1_proof_3} and the above inequality we obtain
$e^{\frac{\sum_{k=1}^{T_2} |y_k-y'_k|}{b_0}} \leq e^{\varepsilon}$ and
\begin{align*}
&\mathbb{P}({\mathcal{M}_1(y_{[1:T_1]})\in D| u_{[0:T_2-1]}, w_{[T_1+1:T_2]}})\cr
\leq& e^{\varepsilon}
\mathbb{P}({\mathcal{M}_1(y'_{[1:T_1]})\in D| u_{[0:T_2-1]}, w_{[T_1+1:T_2]}}).
\end{align*}
Therefore,
\begin{align*}
&\mathbb{P}(\mathcal{M}_1(y_{[1:T_1]})\in D)\notag \\
=&
\int_{\Omega}
\mathbb{P}(\mathcal{M}_1(y_{[1:T_1]})\in D|u_{[0:T_2-1]}, w_{[T_1+1:T_2]})\cr
&\quad \quad \mathrm{d} \mathrm{P}_{u_{[0:T_2-1]}}
\mathrm{d} \mathrm{P}_{w_{[T_1+1:T_2]}}
\notag \\
\leq&
e^{\varepsilon} \int_{\Omega}
\mathbb{P}(\mathcal{M}_1(y'_{[1:T_1]})\in D|u_{[0:T_2-1]}, w_{[T_1+1:T_2]})\cr
&\quad\quad\mathrm{d} \mathrm{P}_{u_{[0:T_2-1]}}
\mathrm{d} \mathrm{P}_{w_{[T_1+1:T_2]}}
=e^{\varepsilon} \mathbb{P}(\mathcal{M}_1(y'_{[1:T_1]})\in D),
\end{align*}
where $\Omega$ is the sample space of $u_{[0:T_2-1]}$ and $w_{[T_1+1:T_2]}$,
$\mathrm{P}_{u_{[0:T_2-1]}}$ and $\mathrm{P}_{w_{[T_1+1:T_2]}}$ are
the probability measures of $u_{[0:T_2-1]}$ and
$w_{[T_1+1:T_2]}$, respectively.  \hfill $\blacksquare$

\begin{remark}
Note that for any $\varepsilon > 0$ and $\delta > 0$, Participant
$\mathcal{P}_0$  can always find $b_0$ sufficiently large to satisfy
\eqref{eq:DP_LS_thm_1_b0}. This implies that Participant $\mathcal{P}_0$
can achieve any privacy protection level by adding sufficiently `large'
noises, no matter how many other participants collude with,
how those participants choose their inputs and what the system noises are.
\end{remark}
\begin{remark}
If $p=0$ in System \eqref{eq:DP_LS_system}, then we can choose $c_0=0$
in \eqref{eq:DP_LS_c0}. Then, we have $C_1 = 1$ in \eqref{eq:DP_LS_C1} and
$\frac{\delta}{b_0}\leq \varepsilon$ in \eqref{eq:DP_LS_thm_1_b0},
which simplifies the selection of $b_0$ to achieve $\varepsilon$-differential
privacy on the system output.
\end{remark}

To establish the differential privacy for Participant~$\mathcal{P}_{i}, i=1, \cdots, m$, the following lemma is needed.

\begin{lemma}
\label{lem:DP_LS_lem_2}
For System \eqref{eq:DP_LS_system},
suppose that all noises and inputs but $u_{i,k_0}$ are zero for some
$i \in  \{1,\cdots,m\}$ and some $k_0\geq 0$.
If $|u_{i,k_0}|\leq \delta$ for some $\delta > 0$ and
Assumption~\ref{ass:DP_LS_A1} holds, then for any $T>0$,
$\sum_{k=1}^T |y_k| \leq C_{i,2} \delta$,
where
\begin{align}
C_{i,2} = (1+\frac{\sqrt{p}c_0\lambda}{1-\lambda})\sum_{j=1}^{q_i}|b_{i,j}|,
\label{eq:DP_LS_C_i2}
\end{align}
with $c_0$ and $\lambda$ given in \eqref{eq:DP_LS_c0}
\end{lemma}
\noindent
{\bf Proof}.
Under the condition of the lemma, System \eqref{eq:DP_LS_system} is equivalent to
\begin{align*}
y_{k+1}&=a_1y_k+a_2y_{k-1}+\cdots+a_py_{k+1-p}
\notag \\
&\quad \,
+ b_{i,1}u_{i,k}
+ \ldots + b_{i,q_{i}}u_{i,k+1-q_{i}},\notag
\end{align*}
where $y_k = 0, \, \forall k\leq k_0,\, u_{i,k}=0, k\neq k_0.$

For $1\leq j \leq p$, let $u_k(j) =
\begin{cases}
 b_{i,j}u_{i,k_0}, & \text{if}\, k=k_0+j-1,\\
 0,   & \text{otherwise}.
\end{cases}$
Then, the above system is equivalent to
\begin{align}
y_{k+1} = a_1y_k+a_2y_{k-1}+\cdots+a_py_{k+1-p}
\notag \\
+ u_k(1) + u_k(2) + \cdots + u_k(q_{i}).
\label{eq:DP_LS_lem_2_proof_1}
\end{align}
Next, consider that $y_{k+1}(j) = a_1y_k(j)+a_2y_{k-1}(j)+\cdots+a_py_{k+1-p}(j)+u_k(j)$
with $y_k(j)=0$ for $k\leq k_0+j$ and $1\leq j\leq q_{i}$.

Similar to \eqref{eq:DP_LS_thm_1_proof_3}, we have
\begin{align*}
\sum_{k=1}^T|y_k(j)| &= \sum_{k=1}^{k_0+j} |y_k(j)|
+ \sum_{k=k_0+j+1}^{T}|y_k(j)|
\notag \\
&\leq
|b_{i,j}|\delta + \sum_{k=k_0+j+1}^T \|y_{[k+1-p:k]}(j)\|_1
\notag \\
&\leq
(1+\frac{\sqrt{p}c_0\lambda}{1-\lambda}) |b_{i,j}|\delta,
\end{align*}
where the last inequality holds for $u_k(j)=0$ when $k\geq k_0+j$.

Therefore, it follows from the linearity of \eqref{eq:DP_LS_lem_2_proof_1} that
\begin{align*}
\sum_{k=1}^T |y_k| =& \sum_{k=1}^T |\sum_{j=1}^{q_{i}}y_k(j)|
=\sum_{j=1}^{q_{i}} \sum_{k=1}^T |y_k(j)|
\notag \\
&\leq (1+\frac{\sqrt{p}c_0\lambda}{1-\lambda})
\sum_{j=1}^{q_{i}}|b_{i,j}|\delta,
\end{align*}
which yields the result. \hfill $\blacksquare$

\begin{theorem}(Differential privacy for Participant~$\mathcal{P}_{i}, i=1, \cdots, m$)
\label{thm:DP_LS_thm_2}
For System \eqref{eq:DP_LS_system} and Algorithm
\eqref{eq:DP_LS_alg_theta}-\eqref{eq:DP_LS_alg_P}, let
$\delta > 0$ and $\varepsilon > 0$ be any prescribed privacy indexes.
If Assumptions~\ref{ass:DP_LS_A1} and \ref{ass:DP_LS_A2} hold, and
Participants $\mathcal{P}_0$ and $\mathcal{P}_i$ choose $b_0$ and $b_i$ such that
\begin{align}
(\frac{C_{i,2}}{b_0}+\frac{1}{b_i}) \delta \leq \varepsilon
\label{eq:DP_LS_thm_1_b_i2}
\end{align}
where $C_{i,2}$ is defined in \eqref{eq:DP_LS_C_i2}, then Algorithm
\eqref{eq:DP_LS_alg_theta}-\eqref{eq:DP_LS_alg_P} is
$\varepsilon$-differentially private under $\delta$-adjacency for
Participant~$\mathcal{P}_{i}$.
\end{theorem}

\noindent
{\bf Proof}. Note that the attacker is able to infer $u_{i,[0:T_1-1]}$ easily according to System \eqref{eq:DP_LS_system}
by colluding with all other participants when the system noises are zero. This makes it impossible to protect Participant $\mathcal{P}_i$'s
sensitive information. In this case, we should also protect Participant $\mathcal{P}_0$'s sensitive information. Then, the attacker's
observed information is $[\bar{y}_{[1:T_2]}^T, \bar{u}_{i,[0:T_2-1]}^T]$, where $T_2 \geq T_1$. We denote this correspondence as
$\mathcal{M}_2(u_{i,[0:T_1-1]}) =
[\bar{y}_{[1:T_2]}^T, \bar{u}_{i,[0:T_2-1]}^T]^T$.
It suffices to show that for any $T_2 \geq T_1$, any $u_{i,[0:T_1-1]}$
and $u'_{i,[0:T_1-1]}$ satisfying $\|u_{i,[0:T_1-1]}-u'_{i,[0:T_1-1]}\|_1
\leq \delta$, and any $D\in \mathscr{B}(\mathbb{R}^{2T_2})$, it holds that
$
\mathbb{P}(\mathcal{M}_2(u_{i,[0:T_1-1]})\in D)\leq e^{\varepsilon}
\mathbb{P}(\mathcal{M}_2(u'_{i,[0:T_1-1]})\in D).
$

Denote
\begin{align*}
u_{-i,[0:T_2-1]} &= [u_{1,[0:T_2-1]}^T,\cdots, u_{i-1,[0:T_2-1]}^T, \\
 &\quad\,\, u_{i+1,[0:T_2-1]}^T,\cdots, u_{m,[0:T_2-1]}^T]^T.
\end{align*}
Note that $\mathbb{P}(\mathcal{M}_2(u_{i,[0:T_1\!-\!1]}\!)\!\!\in \!D)\!=\!
\mathbb{P}([\bar{y}_{[1:T_2]}^T, \bar{u}_{i,[0:T_2\!-\!1]}^T\!]^T \!\!\in
\!D|u_{i,[0:T_1-1]})$
and $y_{[1:T_2]}$ is determined by $u_{-i,[0:T_2-1]}$,
$u_{i,[0:T_1-1]}$, $u_{i,[T_1:T_2-1]}$ and $w_{[1:T_2]}$. Then,
by Assumption~\ref{ass:DP_LS_A2},
we have
\begin{align}
&\!\!\!\!\!\!\!\!\mathbb{P}(\mathcal{M}_2(u_{i,[0:T_1-1]})\!\!\in D|u_{i,[T_1:T_2-1]},
u_{-i,[0:T_2-1]},w_{[1:T_2]})
\notag \\
=&
\int_{\mathbb{R}^{2T_2}} (\frac{1}{2b_0})^{T_2}(\frac{1}{2b_i})^{T_2}
\prod_{k=1}^{T_2}
1_{D|_{\bar{y}_k}}(y_k+\eta_k)e^{-\frac{|\eta_k|}{b_0}}\cr
&\quad \quad   1_{D|_{\bar{u}_{i,k-1}}}(u_{i,k-1}+\xi_{i,k-1})e^{-\frac{|\xi_{i,k-1}|}{b_i}}
\mathrm{d}\Lambda_{T_2} \mathrm{d}\Xi_{T_2}
\notag \\
=&
\int_{\mathbb{R}^{2T_2}} (\frac{1}{2b_0})^{T_2}(\frac{1}{2b_i})^{T_2}
\prod_{k=1}^{T_2}
1_{D|_{\bar{y}_k}}(\bar{y}_k)e^{-\frac{|\bar{y}_k-y_k|}{b_0}}\cr
&\quad \quad   1_{D|_{\bar{u}_{i,k-1}}}(\bar{u}_{i,k-1})e^{-\frac{|\bar{u}_{i,k-1}-u_{i,k-1}|}{b_i}}
\mathrm{d}\bar{\Lambda}_{T_2} \mathrm{d}\bar{\Xi}_{T_2},
\label{eq:DP_LS_thm_2_proof_1}
\end{align}
where
$\mathrm{d}\Lambda_{T_2} = \mathrm{d}\eta_1\cdots\mathrm{d}\eta_{T_2},$$
\mathrm{d}\Xi_{T_2} = \mathrm{d}\xi_{i,0}\cdots\mathrm{d}\xi_{i,T_2-1},$\\$
\mathrm{d}\bar{\Lambda}_{T_2}= \mathrm{d}\bar{y}_{1}\cdots\mathrm{d}\bar{y}_{T_2},$$
\mathrm{d}\bar{\Xi}_{T_2} = \mathrm{d}\bar{u}_{i,0}\cdots\mathrm{d}\bar{u}_{i,T_2-1}.$

Similarly,
\begin{align}
&\!\! \mathbb{P}(\mathcal{M}_2(u'_{i,[0:T_1-1]})\!\!\in D|u_{i,[T_1:T_2-1]},
u_{-i,[0:T_2-1]},w_{[1:T_2]})
\notag \\
&=
\int_{\mathbb{R}^{2T_2}} (\frac{1}{2b_0})^{T_2}(\frac{1}{2b_i})^{T_2}
\prod_{k=1}^{T_2}
1_{D|_{\bar{y}_k}}(\bar{y}_k)e^{-\frac{|\bar{y}_k-y'_k|}{b_0}}\cr
&\quad \quad \quad 1_{D|_{\bar{u}_{i,k-1}}}(\bar{u}_{i,k-1})e^{-\frac{|\bar{u}_{i,k-1}-u'_{i,k-1}|}{b_i}}
\mathrm{d}\bar{\Lambda}_{T_2} \mathrm{d}\bar{\Xi}_{T_2}.
\label{eq:DP_LS_thm_2_proof_2}
\end{align}
When $u_{i,[T_1:T_2-1]}$, $u_{-i,[0:T_2]}$ and
$w_{[1:T_2]}$ are given in System \eqref{eq:DP_LS_system}, we have
\begin{align}
\sum_{k=1}^{T_2}|u_{i,k-1}-u'_{i,k-1}|
\!=\!\|u_{i,[0:T_2-1]}- u'_{i,[0:T_2-1]}\|_1
\!\leq\! \delta.
\label{eq:DP_LS_thm_2_proof_4}
\end{align}
By Lemma \ref{lem:DP_LS_lem_2}, we have
\begin{align}
\sum_{k=1}^{T_2}|y_k-y'_k|
\leq
C_{i,2}\sum_{k=0}^{T_1-1}|u_{i,k}-u'_{i,k}|
\leq C_{i,2}\delta.
\label{eq:DP_LS_thm_2_proof_5}
\end{align}
Note that
\begin{align*}
&\int_{\mathbb{R}^{2T_2}}
\prod_{k=1}^{T_2}
1_{D|_{\bar{y}_k}}(\bar{y}_k)e^{-\frac{|\bar{y}_k-y_k|}{b_0}}
1_{D|_{\bar{u}_{i,k-1}}}(\bar{u}_{i,k-1})\cr
&\quad \quad \quad e^{-\frac{|\bar{u}_{i,k-1}-u_{i,k-1}|}{b_i}}
\mathrm{d}\bar{\Lambda}_{T_2} \mathrm{d}\bar{\Xi}_{T_2}
\notag \\
&\leq
\int_{\mathbb{R}^{2T_2}}
\prod_{k=1}^{T_2}
1_{D|_{\bar{y}_k}}(\bar{y}_k)e^{-\frac{|\bar{y}_k-y'_k|-|y_k-y'_k|}{b_0}}
1_{D|_{\bar{u}_{i,k-1}}}(\bar{u}_{i,k-1})\cr
&\quad \quad \quad e^{-\frac{|\bar{u}_{i,k-1}-u'_{i,k-1}|-|u_{i,k-1}-u'_{i,k-1}|}{b_i}}
\mathrm{d}\bar{\Lambda}_{T_2} \mathrm{d}\bar{\Xi}_{T_2}
\notag \\
&=
e^{\sum_{k=1}^{T_2}\frac{|y_k-y'_k|}{b_0}+\frac{|u_{i,k-1}-u'_{i,k-1}|}{b_i}}
\int_{\mathbb{R}^{2T_2}}
\prod_{k=1}^{T_2}
1_{D|_{\bar{y}_k}}(\bar{y}_k)\cr
&\quad e^{-\frac{|\bar{y}_k-y'_k|}{b_0}}
1_{D|_{\bar{u}_{i,k-1}}}(\bar{u}_{i,k-1})e^{-\frac{|\bar{u}_{i,k-1}-u'_{i,k-1}|}{b_i}}
\mathrm{d}\bar{\Lambda}_{T_2} \mathrm{d}\bar{\Xi}_{T_2}.
\end{align*}
Then, by \eqref{eq:DP_LS_thm_1_b_i2}-\eqref{eq:DP_LS_thm_2_proof_5} one can get
\begin{align*}
&\!\mathbb{P}(\mathcal{M}_2(u_{i,[0:T_1-1]})\in D|u_{i,[T_1:T_2-1]},
u_{-i,[0:T_2-1]},w_{[1:T_2]})
\notag \\
&\!\leq
\!\!e^{\varepsilon}\mathbb{P}(\!\mathcal{M}_2\!(u'_{i,[0:T_1\!-\!1]}\!)\!\in\! D
|u_{i,[T_1:T_2\!-\!1]},\!u_{-i,[0:T_2\!-\!1]},\!w_{[1:T_2]}\!),
\end{align*}
which implies that
\begin{align*}
&\mathbb{P}(\mathcal{M}_2(u_{i,[0:T_1-1]})\in D) \notag \\
&=
\int_{\Omega} \mathbb{P}\left(\mathcal{M}_{2}(u_{i,[0:T_1-1]})\in D|u_{i,[T_{1}:T_{2}-1]},
u_{-i,[0:T_2-1]}\right.,  \notag \\
&\quad \quad \quad \left.\mathrm{d} w_{[1:T_2]}\right)\mathrm{P}_{u_{i,[T_1:T_2-1]}}
\mathrm{d} \mathrm{P}_{u_{-i,[0:T_2-1]}}\mathrm{d} \mathrm{P}_{w_{[1:T_2]}}
\notag \\
&\leq
e^{\varepsilon}
\int_{\Omega}\mathbb{P}\left(\mathcal{M}_2(u'_{i,[0:T_1-1]})\in D|u_{i,[T_1:T_2-1]},
u_{-i,[0:T_2-1]}\right.,\cr
&\quad \quad \quad \left.\mathrm{d} w_{[1:T_2]}\right)\mathrm{P}_{u_{i,[T_1:T_2-1]}}
\mathrm{d} \mathrm{P}_{u_{-i,[0:T_2-1]}}
\mathrm{d} \mathrm{P}_{w_{[1:T_2]}}
\notag \\
&=
e^{\varepsilon}\mathbb{P}(\mathcal{M}_2(u'_{i,[0:T_1-1]})\in D),
\end{align*}
where $\Omega$ is the sample space of $u_{i,[T_1:T_2-1]}$,
$u_{-i,[0:T_2-1]}$ and $w_{[1:T_2]}$, $\mathrm{P}_{u_{i,[T_1:T_2-1]}}$,
$\mathrm{P}_{u_{-i,[0:T_2-1]}}$ and $\mathrm{P}_{w_{[1:T_2]}}$ are
the probability measures of $u_{i,[T_1:T_2-1]}$, $u_{-i,[0:T_2-1]}$
and $w_{[1:T_2]}$, respectively.   \hfill $\blacksquare$

\begin{remark}
Note that for any $\varepsilon>0$ and $\delta>0$, Participant $\mathcal{P}_i$
and Participant $\mathcal{P}_0$ can always find sufficiently large $b_i$ and
$b_0$ to satisfy \eqref{eq:DP_LS_thm_1_b_i2}. This implies that Participant $\mathcal{P}_i$ can achieve any privacy protection level by cooperating with Participant $\mathcal{P}_0$ and
adding sufficient `large' noises, no matter how many other participants
the attacker colludes with, how those participants choose their inputs
and what the system noises are. Furthermore, it can also be seen from
the proof that $\varepsilon$-differential privacy of $u_{i,[0:T_1-1]}$
still holds even if the attacker knows $u_{i,[T_1:T_2]}$ exactly.
\end{remark}

\begin{remark}
In practice,  from \eqref{eq:DP_LS_thm_1_b_i2} it follows that $\frac{\delta}{b_i}> 0$ and $\varepsilon - \frac{C_{i,2}\delta}{b_0} > 0 $ when $C_{i,2} \neq 0$. If we take the same $\delta$ in Theorems 1-2, then by $\varepsilon - \frac{C_{i,2}\delta}{b_0} > 0 $ we have $b_0 > \frac{C_{i,2}\delta}{\varepsilon}$, which further implies that  $b_0$ derived from Theorem 2 must satisfy that $\frac{C_1\delta}{b_0} < \frac{C_1}{C_{i,2}}\varepsilon$ in Theorem 1. This means that the algorithm must first be $\frac{C_1\varepsilon}{C_{i,2}}$-differentially private for Participant~$\mathcal{P}_0$,
and meanwhile there is still some `budget' ($\varepsilon-\frac{C_{i,2}\delta}{b_0}$) left for Participant~$\mathcal{P}_i$ to achieve $\varepsilon$-differential privacy. This is consistent with the fact that we must protect the sensitive information of Participant $\mathcal{P}_0$ in order to protect the sensitive
information of Participant $\mathcal{P}_i$.
\end{remark}


\begin{remark}
If $p=0$, i.e. no regression terms in the output of System~\eqref{eq:DP_LS_system},
then $C_{i,2}= \sum_{j=1}^{q_i}|b_{i,j}|$ in \eqref{eq:DP_LS_C_i2}
which also simplifies the selection of $b_0$ and $b_i$ to achieve
$\varepsilon$-differential privacy on the system input.
\end{remark}

The following theorem shows that if System \eqref{eq:DP_LS_system} is not asymptotically stable, then $\varepsilon$-differential privacy cannot be established.

\begin{theorem}
\label{thm:DP_LS_thm_3} (Necessity of differential privacy)
For System \eqref{eq:DP_LS_system} and Algorithm
\eqref{eq:DP_LS_alg_theta}-\eqref{eq:DP_LS_alg_P}, suppose that
$\omega_k\equiv0$, Assumption~\ref{ass:DP_LS_A2} holds and System
\eqref{eq:DP_LS_system} is not asymptotically stable.
Let $\varepsilon$ and $\delta$ be any positive numbers.
\begin{itemize}
\item For Participant $\mathcal{P}_0$, if $T_1\geq p$, then there exists
$\delta$-adjacent $y_{[1:T_1]}$ and $y'_{[1:T_1]}$, sufficiently large
$T_2$ and measurable set $D_0\in \mathscr{B}(\mathbb{R}^{T_2})$ such
that
$
\mathbb{P}(\mathcal{M}_1(y_{[1:T_1]})\in D_0)\geq
e^{\varepsilon}\mathbb{P}(\mathcal{M}_1(y'_{[1:T_1]})\in D_0).
$
\item For Participant $\mathcal{P}_i, i=1,\cdots,m$, if $T_1 \geq p $ and $b_{i,1}\neq 0$, then there exists $\delta$-adjacent $u_{i,[0:T_1-1]}$ and
$u'_{i,[0:T_1-1]}$, sufficiently large $T_2$ and measurable set
$D_0 \in \mathscr{B}(\mathbb{R}^{2T_2})$ such that
$
\mathbb{P}(\mathcal{M}_2(u_{i,[0:T_1-1]})\in D_0) \geq
e^{\varepsilon} \mathbb{P}(\mathcal{M}_2(u'_{i,[0:T_1-1]})\in D_0).
$
\end{itemize}
\end{theorem}

\noindent
{\bf Proof}.
Since System \eqref{eq:DP_LS_system} is not asymptotically stable,
there exists $z_0 \neq 0\in \mathbb{C}$ such that $|z_0|\leq 1$ and $\lambda(z_0)=0$. Clearly, $\lambda(\bar{z}_0)=0$. Define $\Delta_k=z_0^{-k}+\bar{z}_0^{-k}$ with $z_0^{-1}=re^{i\beta}$ and $\bar{z}_0^{-1}=re^{-i\beta}$, $r\geq 1$.
Then, $\Delta_k = 2r^{k}\cos (k\beta) \in \mathbb{R}$ by Euler's formula
and $\Delta_{k+1} = a_1\Delta_k+a_2\Delta_{k-1}+\cdots+a_p\Delta_{k+1-p}$
for $k \geq p$ by the property of difference equation.
In addition, for any $\beta\in [0, 2\pi)$, there exist $n_0, k_0\in \mathbb{N}$ such that
$n_0\beta \in (2k_0\pi-\frac{\pi}{2}, 2k_0\pi+\frac{\pi}{2})$.
Let $\alpha_0 = |2k_0\pi - n_0\beta| < \frac{\pi}{2}$. Then, there exist
strictly increasing sequences $\{n_i\}_{i\geq 1}$ and $\{k(n_i)\}_{i\geq1}$
such that $n_in_0\beta \in [2k(n_i)\pi - \alpha_0, 2k(n_i)\pi + \alpha_0]$ and $\cos(n_in_0\beta)\geq \cos(\alpha_0) \triangleq \gamma$, where $k(n_i)$ means that integer $k$ depends on $n_i$. Setting $k_i=n_in_0$, we have
$\Delta_{k_i}=2r^{k_i}\cos(k_i\beta) \geq 2\gamma$ for $i\geq 1$.


1) For any given $y_{[1:T_1]}$, let
$y'_{[1:T_1]}= y_{[1:T_1]} + \frac{v_1}{\|v_1\|_1}\delta$ with
$v_1=[\Delta_1, \Delta_2,\cdots, \Delta_{T_1}]^{T}$. Then,
$\|y_{[1:T_1]}-y'_{[1:T_1]}\|_1 \leq \delta$. Further, when
$u_{[0:T_2-1]}$ is given in System \eqref{eq:DP_LS_system}, we have
$y'_{k_i}-y_{k_i} =\frac{\Delta_{k_i}\delta}{\|v_1\|_1}$ for
$i \geq 1$.

Note that $y_{[T_1+1:T_2]}$ is determined by $y_{[1:T_1]}$  and
$u_{[0:T_2-1]}$ when $\omega_k \equiv 0$. Then, we can choose $D_0$ such that
$D_0|_{\bar{y}_{k_i}}=\{x|x\leq y_{k_i}\}$ and
$D_0|_{\bar{y}_k}=\mathbb{R}$ for $k\neq k_i$. Let $n_1$ be the number
of $k_i$ satisfying $k_i < T_2$. Then, we have
\begin{align*}
&\quad \, \mathbb{P}(\mathcal{M}_1(y'_{[1:T_1]})\in D_0|u_{[0:T_2-1]})
\notag \\
&=
\int_{\mathbb{R}^{n_1}}
(\frac{1}{2b_0})^{n_1} \cdot \notag
\prod_{i=1}^{n_1}1_{D|_{\bar{y}_{k_i}}}(\bar{y}_{k_i})
e^{-\frac{|\bar{y}_{k_i}-y'_{k_i}|}{b_0}} \mathrm{d}\bar{\Lambda}_{n_1}
\notag \\
&=
\prod_{i=1}^{n_1}
\int_{-\infty}^{y_{k_i}}
(\frac{1}{2b_0})
e^{-\frac{|\bar{y}_{k_i}-(y_{k_i}+\frac{\Delta_{k_i}\delta}{\|v_1\|_1})|}{b_0}} \mathrm{d}\bar{y}_{k_i}
\notag \\
&=e^{-\frac{\delta\sum_{i=1}^{n_1}\Delta_{k_i}}{\|v_1\|_1b_0}}
\!\!\!\!\int_{\mathbb{R}^{n_1}}\!\!(\frac{1}{2b_0})^{n_1} \cdot
\!\!\prod_{i=1}^{n_1}1_{D|_{\bar{y}_{k_i}}}\!\!(\bar{y}_{k_i})
e^{-\frac{|\bar{y}_{k_i}-y_{k_i}|}{b_0}}\!\!\mathrm{d}\bar{\Lambda}_{n_1}
\notag \\
&=
e^{-\frac{\delta\sum_{i=1}^{n_1}\Delta_{k_i}}{\|v_1\|_1b_0}}
\mathbb{P}(\mathcal{M}_1(y_{[1:T_1]})\in D_0|u_{[0:T_2-1]}),
\end{align*}
where $\mathrm{d}\bar{\Lambda}_{n_1}=\mathrm{d}\bar{y}_{k_1}\mathrm{d}\bar{y}_{k_2}
\cdots\mathrm{d}\bar{y}_{k_{n_1}}$.

Note that $n_1\to \infty$ as $T_2\to \infty$. Then, we can choose sufficiently large
$T_2$ with $n_1 \geq \frac{\|v_1\|_1b_0\varepsilon}{2\delta\gamma}$. Therefore,
\begin{align*}
&\mathbb{P}(\mathcal{M}_1(y_{[1:T_1]})\in D_0|u_{[0:T_2-1]})\notag \\
=&
e^{\frac{\delta\sum_{i=1}^{n_1}\Delta_{k_i}}{\|v_1\|_1b_0}}
\mathbb{P}(\mathcal{M}_1(y'_{[1:T_1]})\in D_0|u_{[0:T_2-1]})
\notag \\
\geq&
e^{\varepsilon} \mathbb{P}(\mathcal{M}_1(y'_{[1:T_1]})\in D_0|u_{[0:T_2-1]}).
\end{align*}
Integrating $u_{[0:T_2-1]}$ on both sides yields the result.

2) Denote $x_k=y'_k-y_k$ and $\delta_k=u_{i,k}-u'_{i,k}$. Then,
\begin{align}
x_{k+1}=&a_1x_k+a_2x_{k-1}+\cdots+a_px_{k+1-p}\cr
&+b_{i,1}\delta_k+b_{i,2}\delta_{k-1}
+\cdots+b_{i,q_i}\delta_{k+1-q_i}.
\label{eq:DP_LS_thm_3_proof_1}
\end{align}
Since $T_1 \geq p$ and $b_{i,1}\neq 0$, we can always find
$\delta_k, 0\leq k < T_1$ such that
$x_{T_1-p+i}=\Delta_{i}$ for $i=1,\cdots,p$.
In fact, we can choose $\delta_k=0$ for $k< T_1-p$. Then, $x_k=0$ for
$1\leq k\leq T_1-p$. By setting $x_{T_1-p+1}=\Delta_1$, we find
$\delta_{T_1-p}=\Delta_1/b_{i,1}$. By setting $x_{T_1-p+2}=\Delta_2$,
we find $\delta_{T_1-p+1}=(\Delta_2-a_1\Delta_1-b_{i,2}\delta_{T_1-p})/b_{i,1}$.
Following this process, we find all the required $\delta_k$ for
$0\leq k <T_1$.

Therefore, for any given $u_{i,[0:T_1-1]}$, let
$u'_{i,[0:T_1-1]}=u_{i,[0:T_1-1]}+\delta\frac{v_2}{\|v_2\|_1}$ with
$v_2=[\delta_0,\delta_1,\cdots, \delta_{T_1-1}]^T$. Then,
$\|u_{i,[0:T_1-1]}-u'_{i,[0:T_1-1]}\|_1\leq \delta$.

Let $u_{i,[T_1,T_2-1]}$ and $u_{-i,[0:T_2-1]}$ be given. Then,
$\delta_{k}=0$ for $k\geq T_1$ and
$x_k=y'_k-y_k= \frac{\delta\Delta_{k+p-T_1}}{\|v_2\|_1}$ for $k> T_1-p.$
Note that $y_{[T_1+1:T_2]}$ is determined by $y_{[1:T_1]}$ and $u_{[0:T_2-1]}$ when
$\omega_k\equiv 0$. Then, we can choose $D_0$ such that $D_0|_{\bar{y}_{k_i+T_1-p}}=\{x|x<y_{k_i+T_1-p}\}$, $D_0|_{\bar{y}_k} =\mathbb{R}$ for $k\neq k_i+T_1-p$ and $D_0|_{\bar{u}_k}=\mathbb{R}$ for
all $k$. Let $n_2$ be the number of $k_i$ satisfying
$k_i+T_1-p\leq T_2$. Then, we have
\begin{align*}
&\!\!\mathbb{P}(\!\mathcal{M}_2\!(u'_{i,[0:T_1-1]})\!\in\! D_0|u_{i,[T_1:T_2-1]},\!
u_{-i,[0:T_2-1]}\!)
\notag \\
=&
\int_{\mathbb{R}^{n_2}}
(\frac{1}{2b_0})^{n_2}
\prod_{i=1}^{n_2}1_{D|_{\bar{y}_{k_i+T_1-p}}}(\bar{y}_{k_i+T_1-p})\cr
&\quad \quad \quad e^{-\frac{|\bar{y}_{k_i+T_1-p}-y'_{k_i+T_1-p}|}{b_0}}
\mathrm{d}\bar{\Lambda}_{n_2} \notag \\
=&\prod_{i=1}^{n_2}
\!\!\int_{-\infty}^{y_{k_i+T_1-p}}
\!\!(\frac{1}{2b_0})^{n_2}
e^{-\frac{|\bar{y}_{k_i\!+\!T_1\!-\!p}\!-\!y_{k_i\!+\!T_1\!-\!p}\!-\!\frac{\delta\Delta_{k_i}}{\|v_2\|_1}|}{b_0}}
\!\!\mathrm{d}\bar{y}_{k_i\!+\!T_1\!-\!p}
\notag \\
=&
e^{\frac{-\delta\sum_{i=1}^{n_2}\Delta_{k_i}}{\|v_2\|_1b_0}}
\int_{\mathbb{R}^{n_2}}
(\frac{1}{2b_0})^{n_2}
\prod_{i=1}^{n_2}1_{D|_{\bar{y}_{k_i+T_1-p}}}(\bar{y}_{k_i+T_1-p})\cr
&\quad \quad \quad e^{-\frac{|\bar{y}_{k_i+T_1-p}-y_{k_i+T_1-p}|}{b_0}}
\mathrm{d}\bar{\Lambda}_{n_2}
\notag \\
=&
e\!^{\frac{-\delta\!\sum_{i=1}^{n_2}\!\!\Delta_{k_i}}{\|v_2\|_1\!b_0}}
\!\mathbb{P}(\!\mathcal{M}_2(u_{i,[0:T_1\!-\!1]}\!)\!\!\in\!\!D_0|u_{i,[T_1:T_2\!-\!1]},\! u_{\!-i,[0:T_2\!-\!1]}),
\end{align*}
where $\mathrm{d}\bar{\Lambda}_{n_2} =\mathrm{d}\bar{y}_{k_1+T_1-p}
\cdots\mathrm{d}\bar{y}_{k_{n_2}+T_1-p}$.

Note that $n_2\to \infty $ as $T_2\to\infty$. Then, we can choose sufficiently large $T_2$ with $ n_2 \geq \frac{\|v_2\|_1b_0\varepsilon}{2\delta\gamma}$. Therefore,
\begin{align*}
&\mathbb{P}(\mathcal{M}_2(u_{i,[0:T_1-1]})\in D_0|u_{i,[T_1:T_2-1]},
u_{-i,[0:T_2-1]})
\notag\\
\geq&e^{\varepsilon}\mathbb{P}(\mathcal{M}_2(u'_{i,[0:T_1-1]})\in
D_0|u_{i,[T_1:T_2-\!1]},u_{-i,[0:T_2-1]}).
\end{align*}
Integrating $u_{i,[T_1:T_2-1]}$, $u_{-i,[0:T_2-1]}$ on both sides yields the result. \hfill $\blacksquare$

\begin{remark}
Theorem \ref{thm:DP_LS_thm_3}  means that if System \eqref{eq:DP_LS_system} is not asymptotically stable, for two adjacent system outputs or system inputs,  as long as the attack duration is sufficiently long, then they can always be distinguished with a high probability, thus the $\varepsilon$-differential privacy cannot be achieved.
\end{remark}

Theorem \ref{thm:DP_LS_thm_3} shows that the ratio of
$\mathbb{P}(\mathcal{M}_1(y_{[1:T_1]})\in D_0)$ and
$\mathbb{P}(\mathcal{M}_1(y'_{[1:T_1]})\in D_0)$ can be arbitrarily large.
However, this does not imply that the attacker can choose $D_0$ to
tell $y_{i,[0:T_1-1]}$ from $y'_{i,[0:T_1-1]}$ easily.
Because $\mathbb{P}(\mathcal{M}_1(y_{[1:T_1]})\in D_0)$ and
$\mathbb{P}(\mathcal{M}_1(y'_{[1:T_1]})\in D_0)$ both can be so small
that the events $\{\mathcal{M}_1(y_{[1:T_1]})\in D_0\}$ and
$\{\mathcal{M}_1(y'_{[1:T_1]})\in D_0\}$ are both almost impossible to
happen in practice. The same reasoning goes to
$\mathbb{P}(\mathcal{M}_2(u_{i,[0:T_1-1]})\in D_0)$ and
$\mathbb{P}(\mathcal{M}_2(u'_{i,[0:T_1-1]})\in D_0)$. This dilemma can be
solved by the following corollary when System \eqref{eq:DP_LS_system}
is unstable.

\begin{corollary}
\label{cor:DP_LS_thm_3_1}
For System \eqref{eq:DP_LS_system} and Algorithm
\eqref{eq:DP_LS_alg_theta}-\eqref{eq:DP_LS_alg_P}, suppose that
$\omega_k\equiv0$, Assumption~\ref{ass:DP_LS_A2} holds and
System \eqref{eq:DP_LS_system} is unstable. Let $\varepsilon$ and
$\delta$ be any positive numbers.
\begin{itemize}
\item For Participant $\mathcal{P}_0$, if $T_1\geq p$, then there exists $\delta$-adjacent $y_{[1:T_1]}$ and $y'_{[1:T_1]}$, sufficiently large $T_2$ and measurable set $ D_0\in \mathscr{B}(\mathbb{R}^{T_2})$ such that  $\mathbb{P}(\mathcal{M}_1(y_{[1:T_1]})\in D_0)
\geq
e^{\varepsilon}\mathbb{P}(\mathcal{M}_1(y'_{[1:T_1]})\in D_0)$
and
$\mathbb{P}(\mathcal{M}_1(y_{[1:T_1]})\in D_0)
=\frac{1}{2}.$
\item For Participant $\mathcal{P}_i, i=1, \cdots, m$, if
$T_1\geq p$ and $b_{i,1}\neq0$, then there exists $\delta$-adjacent
$u_{i,[0:T_1-1]}$ and $u'_{i,[0:T_1-1]}$, sufficiently large $T_2$ and
measurable set $D_0\in \mathscr{B}(\mathbb{R}^{2T_2})$ such that
$\mathbb{P}(\mathcal{M}_2(u_{i,[0:T_1-1]})\in D_0) \geq e^{\varepsilon}
\mathbb{P}(\mathcal{M}_2(u'_{i,[0:T_1-1]})\in D_0)$
and
$\mathbb{P}(\mathcal{M}_2(u_{i,[0:T_1-1]})\in D_0)= \frac{1}{2}.$
\end{itemize}
\end{corollary}
\noindent
{\bf Proof}.
Since System \eqref{eq:DP_LS_system} is unstable, there exists $z_0 \neq0\in \mathbb{C}$
such that $|z_0| < 1$ and $\lambda(z_0)=0$. Define $\Delta_{k}$ as that defined
in the proof of Theorem \ref{thm:DP_LS_thm_3}. Then, there exists
a strictly increasing sequence $\{k_i\}$ and $\gamma > 0$ such that
$\Delta_{k_i}=2r^{k_i}\cos{k_i\beta}\geq 2\gamma r^{k_i}$ with $r > 1$.

1) Since $r>1$, there exists $n_1>0$ such that $\Delta_{k_{n_1}} \geq  \frac{\|v_{1}\|_{1}b_0\varepsilon}{\delta}$. For the $\delta$-adjacent $y_{[1:T_1]}$ and $y'_{[1:T_1]}$ in Theorem~\ref{thm:DP_LS_thm_3}, there exist $T_2 > n_1$ and $D_0 \in \mathscr{B}(\mathbb{R}^{T_2})$ such that $D_0|_{\bar{y}_{k_{n_1}}}=\{x|x\leq y_{k_{n_1}}\}$ and $D_0|_{\bar{y}_{k}}=\mathbb{R}$ for $k\neq k_{n_1}$. Note that
\begin{align*}
&\mathbb{P}(\mathcal{M}_1(y_{[1:T_1]})\in D_0|u_{[0:T_2-1]})\notag \\
=&\mathbb{P}(\bar{y}_{k_{n_1}} \leq y_{k_{n_1}})
=
\int_{-\infty}^{y_{k_{n_1}}} \frac{1}{2b_0}
e^{-\frac{|\bar{y}_{k_{n_1}}-y_{k_{n_1}}|}{b_0}}
\mathrm{d}\bar{y}_{k_{n_1}}=\frac{1}{2}.
\end{align*}
Integrating $u_{[0:T_2-1]}$ on both
sides yields $\mathbb{P}(\mathcal{M}_1(y_{[1:T_1]})\in D_0)=\frac{1}{2}$.
Similar to Theorem \ref{thm:DP_LS_thm_3} we can prove
the remaining result.

2) Since $r>1$, there exists $n_2>0$ such that
$\Delta_{k_{n_2}}>\frac{\|v_{2}\|_{1}b_0\varepsilon}{\delta}$. For the $\delta$-adjacent
$u_{i,[0:T_1-1]}$ and $u'_{i,[0:T_1-1]}$ in Theorem \ref{thm:DP_LS_thm_3},
there exist $T_2 > n_2$ and $D_0 \in \mathscr{B}(\mathbb{R}^{2T_2})$
such that $D_0|_{\bar{y}_{k_{n_2}}}=\{x|x\leq y_{k_{n_2}}\}$,
$D_0|_{\bar{y}_{k}}=\mathbb{R}$ for $k\neq k_{n_2}$ and
$D_0|_{\bar{u}_{i,k}}=\mathbb{R}$ for all $k$. Following the reasoning of
the former part yields the result.   \hfill $\blacksquare$

\begin{remark}
If the attacker knows $u_{i,[0:T_2-1]}$, $i=1,\cdots,m$, then from Corollary \ref{cor:DP_LS_thm_3_1}, the attacker can choose a sequence of $\{\varepsilon(l)\}_{l=1}^{\infty}$ with $\varepsilon(l)
\uparrow \infty$. For each $\varepsilon(l)$, there exist $T_2(l) \in
\mathbb{N}$ satisfying $T_2(l) > T_2(l-1)$ and
$D_0(l) \in \mathscr{B}(\mathbb{R}^{T_2(l)})$ such that
$\mathbb{P}(\mathcal{M}_1(y_{[1:T_1]})\in D_0(l)) \geq \frac{1}{2}$ and
$\mathbb{P}(\mathcal{M}_1(y'_{[1:T_1]})\in D_0(l)) \leq e^{-\varepsilon(l)}.$
In addition, for a fixed $T_2$, it can be verified that
$\{\bar{y}_{k_{n_1(l)}} \in D_0(l)\}_l$ are independent for $l$
satisfying $T_2(l) \leq T_2$. Let $T_2 \to \infty$, by Borel-Cantelli
theorem {\rm \cite{chow1988probability}}, $\mathbb{P}(\bar{y}_{k_{n_1(l)}}\in D_0(l), i.o.)=1$, i.e. $\{\bar{y}_{k_{n_1(l)}}\in D_0(l)\}$
happens infinitely many times, and $\mathbb{P}(\bar{y'}_{k_{n_1(l)}}\in D_0(l))\downarrow 0$. In this case, for the $\delta$-adjacent $y_{[1:T_1]}$ and $y'_{[1:T_1]}$, the attacker will
finally observe $\{\bar{y}_{k_{n_1(l)}}\in D_0(l)\}$ and declare that
$\bar{y}_{k_{n_1(l)}}$ is from $y_{[1:T_1]}$ rather than $y'_{[1:T_1]}$
with arbitrarily large probability. The same reasoning works for
$u_{i,[0:T_1-1]}$ and $u'_{i,[0:T_1-1]}$. Therefore, we assert that if
System \eqref{eq:DP_LS_system} is deterministic and unstable, as long
as the attack duration is sufficiently long, the attacker can always
find a sequence of sets $D_0$ to distinguish the sensitive information
from some of its $\delta$-adjacent neighbors.
\end{remark}

\subsection{Performance analysis}
In this subsection, the estimation error of the algorithm will be discussed. To do so, we introduce the following useful assumptions and lemmas.
\begin{assumption} \label{ass:DP_LS_A3}
$\{\omega_k, \mathscr{F}_k\}$ is a martingale difference sequence with $\{\mathscr{F}_k\}$ being a nondecreasing $\sigma$-algebra sequence
and there exists $\beta \geq 2$ such that
$ \sup_{k} E(|\omega_{k+1}|^\beta | \mathscr{F}_k) < \infty,
\quad \text{a.s.} $
\end{assumption}

\begin{remark}
Assumption \ref{ass:DP_LS_A3} is a standard assumption in the asymptotic analysis of system identification of linear systems \cite{lai1982least,Sayedana2024,chen2014recursive} and allows the noise process to be non-stationary.  In Assumption \ref{ass:DP_LS_A3}, a sequence of martingale differences is broader  than a sequence of independent variables, which implies a much weaker restriction on sequence memory than independence and allows $\omega_{k+1}$ to depend on $\mathscr{F}_k$. Many random variables, such as Gaussian random variables, uniformly distributed random variables, satisfy this assumption.
\end{remark}
\begin{assumption}\label{ass:DP_LS_A4}
The regression vector $\{\varphi_k,\mathscr{F}_k\}$ is an adaptive sequence,
i.e. $\varphi_k \in \mathscr{F}_k, \forall k\geq 0$.
\end{assumption}

\begin{remark}
In fact, the $\sigma$-algebra $\mathscr{F}_k$ in Assumptions~\ref{ass:DP_LS_A3} and \ref{ass:DP_LS_A4} can be set to be
$\sigma\{y_l,u_{i,l},\omega_l,\eta_l, \xi_{i,l}|i=1,\cdots,m, l=0,1,\cdots,k\}$
with $y_l=0, \omega_l=0, \eta_l=0$ for $l\leq 0$.
\end{remark}
\begin{lemma}
\label{lem:DP_LS_sum_phi}
For Algorithm \eqref{eq:DP_LS_alg_theta}-\eqref{eq:DP_LS_alg_a}, $\sum_{t=0}^k \bar{a}_t\bar{\varphi}_t^T\bar{P}_t\bar{\varphi}_t
\leq \ln|\bar{P}_{k+1}^{-1}| - \ln|\bar{P}_0^{-1}|.$
\end{lemma}
\noindent
{\bf Proof}. The proof of this lemma is similar to Theorem 3.2.1 of \cite{chen2014recursive}, so we omit it here.   \hfill $\blacksquare$

Following the ideas of classical least-squares methods in linear stochastic regression model \cite{moore1978strong,lai1982least,guo1991Astrom}, we have the following result for further convergence analysis of the algorithm.

\begin{lemma}
\label{lem:DP_LS_lem_4}
For System \eqref{eq:DP_LS_system} and Algorithm \eqref{eq:DP_LS_alg_theta}-\eqref{eq:DP_LS_alg_P}, if Assumptions~\ref{ass:DP_LS_A2}-\ref{ass:DP_LS_A4} hold, then there exist $\kappa>1, \gamma>1$, such that
\begin{align*}
&\tilde{\theta}_{k+1} \bar{P}_{k+1}^{-1} \tilde{\theta}_{k+1}
+ (1-\frac{1}{\gamma}+o(1))\sum_{t=0}^k
\bar{a}_t(\tilde{\theta}_t^T\bar{\varphi}_t)^2
\notag \\
=&\left\{
\begin{array}{lc}
\!\!O(\ln r_k (\ln(e+ \ln r_k))^{\kappa})+ (1+\gamma+o(1))s_k, & \beta=2; \\
\!\!O(\ln r_k)+ (1+\gamma+o(1))s_k, &  \beta>2,
\end{array} {\rm a.s.}\right.
\end{align*}
where $\tilde{\theta}_k = \theta-{\theta}_k,
s_k=\sum_{t=0}^k(\theta^T(\varphi_t-\bar{\varphi}_t))^2$
and
\begin{align}
r_k = e+ \sum_{t=0}^k\|\bar{\varphi}_t\|^2.
\label{eq:DP_LS_rt}
\end{align}
\end{lemma}
\noindent
{\bf Proof}.
Substituting \eqref{eq:DP_LS_system_compressed} and \eqref{eq:DP_LS_ybar}
into \eqref{eq:DP_LS_alg_a} yields
\begin{align*}
\theta_{k+1}
&=
\theta_k + \bar{a}_k\bar{P}_k\bar{\varphi}_k(\theta^T\varphi_k
+\omega_{k+1} + \eta_{k+1} -\theta_k^T\bar{\varphi}_k),
\end{align*}
which implies
\begin{align}
&\tilde{\theta}_{k+1}
=
\theta-\theta_{k+1}
\notag \\
=&
\tilde{\theta}_k-\bar{a}_k\bar{P}_k\bar{\varphi}_k
(\theta^T\varphi_k+\omega_{k+1} + \eta_{k+1} -\theta_k^T\bar{\varphi}_k)
\notag \\
=&
\tilde{\theta}_k-\bar{a}_k\bar{P}_k\bar{\varphi}_k
(\theta^T\varphi_k-\theta^T\bar{\varphi}_k+\theta^T\bar{\varphi}_k
-\theta_k^T\bar{\varphi}_k\cr
&+\omega_{k+1}+\eta_{k+1})
\notag\\
=&\tilde{\theta}_k \!-\! \bar{a}_k\bar{P}_k\bar{\varphi}_k
[\tilde{\theta}_k^T\bar{\varphi}_k \!+\! \theta^T(\varphi_k\!-\!\bar{\varphi}_k)
\!+\!\omega_{k+1}\!+\!\eta_{k+1}]
\notag \\
=&
\tilde{\theta}_k - \bar{a}_k\bar{P}_k\bar{\varphi}_k
(\tilde{\theta}_k^T\bar{\varphi}_k+ d_k + v_{k+1}),
\label{eq:DP_LS_lem_4_proof_1}
\end{align}
where $d_k=\theta^T(\varphi_k-\bar{\varphi}_k)$ and
$v_k=\omega_{k}+\eta_{k}$.

Consider the following Lyapunov function: $V_k=\tilde{\theta}_k^T\bar{P}_{k}^{-1}\tilde{\theta}_k.$
Then, by \eqref{eq:DP_LS_lem_4_proof_1} we have
\begin{align}
&V_{k+1}
=
\tilde{\theta}_{k+1}^T\bar{P}_{k+1}^{-1}\tilde{\theta}_{k+1}
\notag\\
=&
[\tilde{\theta}_k - \bar{a}_k\bar{P}_k\bar{\varphi}_k
(\tilde{\theta}_k^T\bar{\varphi}_k+ d_k + v_{k+1})]^T
\cdot \bar{P}_{k+1}^{-1}
\notag \\
&
\cdot[\tilde{\theta}_k - \bar{a}_k\bar{P}_k\bar{\varphi}_k
(\tilde{\theta}_k^T\bar{\varphi}_k+ d_k + v_{k+1})]
\notag \\
=&
\tilde{\theta}_k^T\bar{P}_{k+1}^{-1}\tilde{\theta}_k
-2 \bar{a}_k\tilde{\theta}_k^T\bar{P}_{k+1}^{-1}\bar{P}_k\bar{\varphi}_k
(\tilde{\theta}_k^T\bar{\varphi}_k+ d_k + v_{k+1})
\notag \\
&+\bar{a}_k^2\bar{\varphi}_k^T\bar{P}_k\bar{P}_{k+1}^{-1}
\bar{P}_k\bar{\varphi}_k
(\tilde{\theta}_k^T\bar{\varphi}_k+ d_k + v_{k+1})^2.
\label{eq:DP_LS_lem_4_proof_2}
\end{align}
For the first term in \eqref{eq:DP_LS_lem_4_proof_2},
by \eqref{eq:DP_LS_alg_P} we have
\begin{align}
\!\!\!\tilde{\theta}_k^T\bar{P}_{k+1}^{-1}\tilde{\theta}_k
\!=\!\tilde{\theta}_k^T
(\bar{P}_k^{-1}\!+\!\bar{\varphi}_k\bar{\varphi}_k^T)
\tilde{\theta}_k
\!=\!\tilde{\theta}_k^T\bar{P}_k^{-1}\tilde{\theta}_k \!+\!
(\tilde{\theta}_k^T\bar{\varphi}_k)^2.
\label{eq:DP_LS_lem_4_proof_3}
\end{align}
For the last two terms in \eqref{eq:DP_LS_lem_4_proof_2},
by \eqref{eq:DP_LS_alg_theta} and \eqref{eq:DP_LS_alg_P} we have
\begin{align*}
&\bar{a}_k\bar{P}_{k+1}^{-1}\bar{P}_k\bar{\varphi}_k
=
\bar{a}_k(I+\bar{\varphi}_k\bar{\varphi}_k^T\bar{P}_k)\bar{\varphi}_k
\notag \\
=&
\bar{a}_k\bar{\varphi}_k(1+\bar{\varphi}_k^T\bar{P}_k\bar{\varphi}_k)
=
\bar{\varphi}_k,
\end{align*}
and $\bar{a}_{k}\bar{\varphi}_k^T\bar{P}_k\bar{\varphi}_k = 1-\bar{a}_k.$

Therefore,
\begin{align}
&\bar{a}_k\tilde{\theta}_k^T\bar{P}_{k+1}^{-1}\bar{P}_k\bar{\varphi}_k
(\tilde{\theta}_k^T\bar{\varphi}_k+ d_k + v_{k+1})
\notag \\
&=
\tilde{\theta}_k^T\bar{\varphi}_k
(\tilde{\theta}_k^T\bar{\varphi}_k+ d_k + v_{k+1})
\notag \\
&=
(\tilde{\theta}_k^T\bar{\varphi}_k)^2
+\tilde{\theta}_k^T\bar{\varphi}_kd_k
+\tilde{\theta}_k^T\bar{\varphi}_kv_{k+1},
\label{eq:DP_LS_lem_4_proof_4}
\end{align}
and
\begin{align}
&
\bar{a}_k^2\bar{\varphi}_k^T\bar{P}_k\bar{P}_{k+1}^{-1}
\bar{P}_k\bar{\varphi}_k
(\tilde{\theta}_k^T\bar{\varphi}_k+ d_k + v_{k+1})^2
\notag \\
&=
\bar{a}_k\bar{\varphi}_k^T\bar{P}_{k}\bar{\varphi}_k
(\tilde{\theta}_k^T\bar{\varphi}_k+ d_k + v_{k+1})^2
\notag \\
&=
(1-\bar{a}_k) (\tilde{\theta}_k^T\bar{\varphi}_k+ d_k + v_{k+1})^2.
\label{eq:DP_LS_lem_4_proof_5}
\end{align}
By substituting \eqref{eq:DP_LS_lem_4_proof_3}-\eqref{eq:DP_LS_lem_4_proof_5}
into \eqref{eq:DP_LS_lem_4_proof_2}, it follows that
\begin{align*}
V_{k+1}
=&
\tilde{\theta}_k^T\bar{P}_k^{-1}\tilde{\theta}_k +
(\tilde{\theta}_k^T\bar{\varphi}_k)^2
-2[(\tilde{\theta}_k^T\bar{\varphi}_k)^2
+\tilde{\theta}_k^T\bar{\varphi}_kd_k
\notag \\
&+\tilde{\theta}_k^T\bar{\varphi}_kv_{k+1}]+(1-\bar{a}_k) (\tilde{\theta}_k^T\bar{\varphi}_k+ d_k + v_{k+1})^2
\notag \\
=&
\tilde{\theta}_k^T\bar{P}_k^{-1}\tilde{\theta}_k
+[1-2+(1-\bar{a}_k)] (\tilde{\theta}_k^T\bar{\varphi}_k)^2
\notag \\
&+\![-2\!+\!2(1\!-\!\bar{a}_k)]\tilde{\theta}_k^T\bar{\varphi}_kd_k\!+\![-2\!+\!2(1\!-\!\bar{a}_k)] \tilde{\theta}_k^T\bar{\varphi}_kv_{k\!+\!1}\notag \\
&+ (1-\bar{a}_k)d_k^2 + 2(1-\bar{a}_k)d_kv_{k+1}
+ (1-\bar{a}_k)v_{k+1}^2
\notag \\
=&
\tilde{\theta}_k^T\bar{P}_k^{-1}\tilde{\theta}_k
\!-\!\bar{a}_k(\tilde{\theta}_k^T\bar{\varphi}_k)^2
\!-\!2\bar{a}_k \tilde{\theta}_k^T\bar{\varphi}_kd_k
\!-\!2\bar{a}_k \tilde{\theta}_k^T\bar{\varphi}_kv_{k+1}
\notag \\
&+ (1-\bar{a}_k)d_k^2 + 2(1-\bar{a}_k)d_kv_{k+1}
+ (1-\bar{a}_k)v_{k+1}^2.
\end{align*}
Then, summing up the above equations from time 0 to time $k$ leads to
\begin{align}
&\!\!V_{k+1} + \sum_{t=0}^k \bar{a}_t(\tilde{\theta}_t^T\bar{\varphi}_t)^2\cr
\!\!\!\!\!=\!&
V_0 \!- \!2 \sum_{t=0}^k \bar{a}_t\tilde{\theta}_t^T\bar{\varphi}_td_t
\!-\!2 \sum_{t=0}^k \bar{a}_t\tilde{\theta}_t^T\bar{\varphi}_tv_{t+1}\!+\!\sum_{t=0}^k(1-\bar{a}_t)d_t^2
\notag \\
&\!\!+ 2\sum_{t=0}^k(1-\bar{a}_t)d_tv_{t+1}+ \sum_{t=0}^k(1-\bar{a}_t)v_{t+1}^2.
\label{eq:DP_LS_lem_4_proof_6}
\end{align}
Now, we deal with summation terms on the right-hand side of
\eqref{eq:DP_LS_lem_4_proof_6} one by one.

1)  Note that
$\forall a,b \in \mathbb{R}, \gamma > 1, 2ab \leq \gamma a^2+ \frac{1}{\gamma}b^2$.
Then,
\begin{align}
&-2\sum_{t=0}^k \bar{a}_t\tilde{\theta}_t^T\bar{\varphi}_td_t
\leq
\sum_{t=0}^k 2|\bar{a}_t\tilde{\theta}_t^T\bar{\varphi}_td_t| \notag \\
\leq&
\sum_{t=0}^k  [\gamma d_t^2
+\frac{1}{\gamma}(\bar{a}_t\tilde{\theta}_t^T\bar{\varphi}_t )^2]\leq\sum_{t=0}^k \{ \gamma d_t^2
+\frac{1}{\gamma}\bar{a}_t(\tilde{\theta}_t^T\bar{\varphi}_t )^2\}
\notag \\
=&\gamma s_k +
\frac{1}{\gamma}\sum_{t=0}^k\bar{a}_t(\tilde{\theta}_t^T\bar{\varphi}_t )^2.
\label{eq:DP_LS_lem_4_proof_7}
\end{align}
2) Since $\eta_k \sim L(0,b_0)$, $E|\eta_k|^{x} < \infty$ for all $ k > 0$ and any $x \geq 0$. Consequently, by Assumption~\ref{ass:DP_LS_A3} and $C_r$-inequality
\cite{chen2014recursive},
\begin{align}
\!\!\!\!&\sup_k E(|v_{k+1}|^{\beta}|\mathscr{F}_{k})\cr
\!\!\!\!\leq& 2^{\beta-1}\sup_k( E(|\omega_{k+1}|^{\beta}|\mathscr{F}_k)
\!+\!E(|\eta_{k+1}|^{\beta}|\mathscr{F}_k))\!<\!\infty,
\label{eq:DP_LS_lem_4_proof_14}
\end{align}
which implies $\sup_k E(|v_{k+1}|^{2}|\mathscr{F}_{k}) < \infty$.

Note that $\bar{a}_k\tilde{\theta}_k^T\bar{\varphi}_k\in \mathscr{F}_k$.
Then, by Theorem 1.2.14 in \cite{chen2014recursive},
there exist $\delta_1 > 0$ and $\delta_2 \in (0,\frac{1}{2})$ such that
\begin{align}
&\quad \,\sum_{t=0}^k\bar{a}_t\tilde{\theta}_t^T\bar{\varphi}_tv_{t+1}
\notag \\
&=
O( [ \sum_{t=0}^k ( \bar{a}_t\tilde{\theta}_t^T\bar{\varphi}_t )^2 ]
^{\frac{1}{2}}(\ln (e + \sum_{t=0}^k (\bar{a}_t\tilde{\theta}_t^T
\bar{\varphi}_t)^2 ))^{\frac{1}{2}+\delta_1})
\notag \\
&=
O( [ \sum_{t=0}^k ( \bar{a}_t\tilde{\theta}_t^T\bar{\varphi}_t )^2 ]
^{\frac{1}{2}}(O(1)+
o([\sum_{t=0}^k(\bar{a}_t\tilde{\theta}_t^T\bar{\varphi}_t )^2]^{\delta_2})))
\notag \\
&=
O(1) + o(\sum_{t=0}^k \bar{a}_t(\tilde{\theta}_t^{T}\varphi_t)^2)
\quad \text{a.s. } 
\label{eq:DP_LS_lem_4_proof_8}
\end{align}
where the second equation comes from $\lim_{x\to\infty}
\frac{(\ln(e+x))^{\frac{1}{2}+\delta_1}}{x^{\delta_2}}$ $=0$.

3) Since $0\leq \bar{a}_k\leq 1$, by \eqref{eq:DP_LS_alg_theta}, we have
\begin{align}
\sum_{t=0}^k (1-\bar{a}_t)d_t^2 \leq \sum_{t=0}^k d_t^2 = s_k.
\label{eq:DP_LS_lem_4_proof_9}
\end{align}
4) Note that $(1-\bar{a}_k)d_k\in\mathscr{F}_k$,
$\sup_k(|v_{k+1}|^2|\mathscr{F}_k) < \infty$ and
$\sum_{t=0}^k d_t^2 = s_k$. Then, by
Theorem 1.2.14 in \cite{chen2014recursive},
there exist $\delta_3 > 0$ and $\delta_4 \in (0, \frac{1}{2})$ such that
\begin{align}
&\quad \, \sum_{t=0}^k (1-\bar{a}_t)d_tv_{t+1}
\notag \\
&=
O([\sum_{t=0}^k((1-\bar{a}_t)d_t)^2]^{\frac{1}{2}}
[\ln(e+\sum_{t=0}^k((1-\bar{a}_t)d_t)^2)]^{\frac{1}{2}+\delta_3})
\notag\\
&=
O(s_k^{\frac{1}{2}}(\ln(e+s_k))^{\frac{1}{2}+\delta_3})
=
O(s_k^\frac{1}{2}(O(1)+o(s_k^{\delta_4})))
\notag \\
&=
O(1) + o(s_k)
\quad \text{a.s.}\, 
\label{eq:DP_LS_lem_4_proof_10}
\end{align}
5) Note that by \eqref{eq:DP_LS_alg_P},
$\bar{P}_{k+1}^{-1}=\bar{P}_0^{-1}
+\sum_{t=0}^k\bar{\varphi}_t\bar{\varphi}_t^T$ is positive definite.
Then,
\begin{align*}
|\bar{P}_{k+1}^{-1}|
\leq&\lambda_{\max}^{p+\sum_{i=1}^m q_i}
(\bar{P}_{k+1}^{-1})
< (tr{\bar{P}_{k+1}})^{p+\sum_{i=1}^m q_i}\cr
=&(tr{\bar{P}_0^{-1}}+\sum_{t=0}^k\|\bar{\varphi}_t\|^2)^{p+\sum_{i=1}^m q_i}.
\end{align*}
Consequently, by \eqref{eq:DP_LS_rt} one can get
\begin{align}
\ln|\bar{P}_{k+1}^{-1}|=O(1)+O(\ln r_k).
\label{eq:DP_LS_lem_4_proof_11}
\end{align}
By using $C_r$-inequality and Lyapunov inequality \cite{chen2014recursive},
it follows from \eqref{eq:DP_LS_lem_4_proof_14} that for any
$\alpha\in [2, \min\{\beta, 4\}]$,
\begin{align*}
&\quad \, \sup_k E[|v_{k+1}^2 -  E(v_{k+1}^2|\mathscr{F}_k)|^{\frac{\alpha}{2}}|
\mathscr{F}_k]
\notag \\
&\leq
\sup_{k} 2^{\frac{\alpha}{2}-1} ( E[|v_{k+1}^2|^{
\frac{\alpha}{2}}|\mathscr{F}_k]
+ E[| E(v_{k+1}^2|\mathscr{F}_k)|^{\frac{\alpha}{2}}
|\mathscr{F}_k])
\notag \\
&\leq
\sup_{k} 2^{2-1}( E[|v_{k+1}^2|^{\frac{\alpha}{2}}|\mathscr{F}_k]
+| E(v_{k+1}^2|\mathscr{F}_k)|^{\frac{\alpha}{2}})
\notag \\
&\leq
2\sup_{k}( E[|\omega_{k+1}|^{\alpha}|\mathscr{F}_k]
+( E[|\omega_{k+1}|^{\alpha}|\mathscr{F}_k])^{
\frac{1}{\alpha}\cdot \alpha})
\notag \\
&=
4\sup_{k} E[|v_{k+1}|^{\alpha}|\mathscr{F}_k] < \infty,
\quad \text{a.s. }
\end{align*}
Note that $1-\bar{a}_k=\bar{a}_k\bar{\varphi}_k^T\bar{P}_k\bar{\varphi}_k
\in \mathscr{F}_k$ by \eqref{eq:DP_LS_alg_P} and
$\{v_{k+1}^2 -  E(v_{k+1}^2|\mathscr{F}_k)\}$ is a martingale difference
sequence with respect to $\{\mathscr{F}_k\}$.
It follows from Theorem 1.2.14 in \cite{chen2014recursive},
Lemma \ref{lem:DP_LS_sum_phi} and \eqref{eq:DP_LS_lem_4_proof_11} that for any $\delta_5 > 0$ such that
\begin{align*}
&
\sum_{t=0}^k (1-\bar{a}_t)
(v_{t+1}^2- E[v_{t+1}^2|\mathscr{F}_t])
\notag \\
=&
O([\sum_{t=0}^k(1-\bar{a}_t)^{\frac{\alpha}{2}}]^{\frac{2}{\alpha}}
[\ln(1 +\sum_{t=0}^k(1-\bar{a}_t)^{\frac{\alpha}{2}})]^{\frac{2}{\alpha}+\delta_5})
\notag \\
=&
O([\sum_{t=0}^k(1-\bar{a}_t)]^{\frac{2}{\alpha}}[\ln(1 +\sum_{t=0}^k
(1-\bar{a}_t))]^{\frac{2}{\alpha}+\delta_5})
\notag \\
=&
O((\ln|\bar{P}_{k+1}^{-1}|-\ln|\bar{P}_0^{-1}|)^{\frac{2}{\alpha}}
(\ln(1 +\ln|\bar{P}_{k+1}^{-1}|\notag \\
&-\ln|\bar{P}_0^{-1}|))^{\frac{2}{\alpha}
+\delta_5})
\notag \\
=&
O(1) + O((\ln r_k)^{\frac{2}{\alpha}}
(\ln(e+\ln r_k))^{\frac{2}{\alpha}+\delta_5})
\quad \text{a.s. }
\end{align*}
For the last term in the above equation, if $\beta=2$, then
$\frac{2}{\alpha}=1$; else if $\beta>2$, then $\alpha$ can be taken
greater than 2 with $\frac{2}{\alpha}<1$ and there exists $\delta_6\in (0,1-\frac{2}{\alpha})$ such that
\begin{align*}
&O((\ln r_k)^{\frac{2}{\alpha}}
(\ln(e+\ln r_k))^{\frac{2}{\alpha}+\delta_5})\cr
=& O((\ln r_k)^{\frac{2}{\alpha}}(O(1)+o((\ln r_k)^{\delta_6})))
= O(1)+O(\ln r_k).
\end{align*}
Thus, for both cases, set $\kappa=1+\delta_5>1$, we always have
\begin{align*}
&\quad \,
\sum_{t=0}^k (1-\bar{a}_t)
(v_{t+1}^2- E[v_{t+1}^2|\mathscr{F}_t])
\notag \\
&=O(1)+O(\ln r_k (\ln(e+ \ln r_k))^{\kappa 1_{\{\beta=2\}}})
\quad \text{a.s.} 
\end{align*}
Therefore, by Assumption~\ref{ass:DP_LS_A3}, Lemma \ref{lem:DP_LS_sum_phi}, \eqref{eq:DP_LS_lem_4_proof_11}
and the above equation,
\begin{align}
&\sum_{t=0}^k (1-\bar{a}_t)v_{t+1}^2
\notag \\
=&
\sum_{t=0}^k (1-\bar{a}_t)(v_{t+1}^2 - E(v_{t+1}^2|\mathscr{F}_t))\notag \\
&+\sum_{t=0}^k  (1-\bar{a}_t) E(v_{t+1}^2|\mathscr{F}_t)
\notag \\
\leq&
\sum_{t=0}^k (1-\bar{a}_t)(v_{t+1}^2 - E(v_{t+1}^2|\mathscr{F}_t))
+\sigma \sum_{t=0}^k (1-\bar{a}_t)
\notag \\
=&
O(1) + O(\ln r_k (\ln(1 +\ln r_k))^{\kappa 1_{\{\beta=2\}}})
+ O(\ln r_k)
\notag \\
=&
O(1) + O(\ln r_k (\ln(e+ \ln r_k))^{\kappa 1_{\{\beta=2\}}})
\quad \text{a.s.},
\label{eq:DP_LS_lem_4_proof_12}
\end{align}
where $\sigma = \sup_k  E(v_{k+1}^2|\mathscr{F}_k)$ by Assumption~\ref{ass:DP_LS_A3}.

Substituting \eqref{eq:DP_LS_lem_4_proof_7}, \eqref{eq:DP_LS_lem_4_proof_8}-\eqref{eq:DP_LS_lem_4_proof_10}
and \eqref{eq:DP_LS_lem_4_proof_12} into \eqref{eq:DP_LS_lem_4_proof_6}, we have
\begin{align*}
&V_{k+1} + (1-\frac{1}{\gamma}+o(1))\sum_{t=0}^k
\bar{a}_t(\tilde{\theta}_t^T\bar{\varphi}_t)^2
\notag \\
= &O(1) \!+\! (1\!+\!\gamma\!+\!o(1))s_k\!+\! O(\ln r_k (\ln(e\!+\! \ln r_k))^{\kappa 1_{\{\beta=2\}}})
\quad \text{a.s. }
\end{align*}
Note that $\ln r_k\geq1$. Then, the result is obtained. \hfill $\blacksquare$

\begin{theorem}
\label{thm:DP_LS_thm_4}
For System \eqref{eq:DP_LS_system} and Algorithm
\eqref{eq:DP_LS_alg_theta}-\eqref{eq:DP_LS_alg_P}, suppose that
Assumptions~\ref{ass:DP_LS_A1}-\ref{ass:DP_LS_A4},
\eqref{eq:DP_LS_thm_1_b0} and \eqref{eq:DP_LS_thm_1_b_i2} hold.
If there exist $\kappa>1$ and $\gamma_1>0$, 
\begin{align}
&\left\{
\begin{array}{lc}
\frac{\ln r_{k-1} (\ln(e+\ln r_{k-1}))^{\kappa}}
{\lambda_{\min}(\bar{P}_{k}^{-1})} \to \!0, & \beta=2; \\
\frac{\ln r_{k-1}}
{\lambda_{\min}(\bar{P}_{k}^{-1})} \to \!0, &  \beta>2, 
\end{array} \right.
 {\rm a.s.}  \quad {\rm as} \ \, k\to\infty, 
\label{ass:DP_LS_thm_4_pe_1} \\
&\gamma_1=\liminf_{k > 0} \lambda_{\min} (\frac{1}{k} \bar{P}_{k}^{-1}) > 0
\quad  {\rm a.s. },
\label{ass:DP_LS_thm_4_pe_2}
\end{align}
then
\renewcommand{\labelenumi}{\roman{enumi}.}
\begin{enumerate}
\item Algorithm
\eqref{eq:DP_LS_alg_theta}-\eqref{eq:DP_LS_alg_P} is $\varepsilon$-differentially private under $\delta$-adjacency for each participant;
\item For $\tilde{\theta}_k = \theta-\theta_k$, we have
\begin{align}
\lim_{k\to\infty} \|\tilde{\theta}_k\|
\leq
2 \|\theta\| \sqrt{\frac{pb_0^2\!+\!\sum_{i=1}^{m}q_ib_i^2}
{\gamma_1}},
\quad \!\! {\rm a.s.}.
\label{eq:DP_LS_thm_4_error_bound}
\end{align}
\end{enumerate}
\end{theorem}

\noindent
{\bf Proof}.
The $\varepsilon$-differential privacy follows from Theorems~\ref{thm:DP_LS_thm_1}, \ref{thm:DP_LS_thm_2}. Next, we only need to prove \eqref{eq:DP_LS_thm_4_error_bound}.

By Lemma \ref{lem:DP_LS_lem_4}, for sufficiently large $k$,
\begin{align}
&\|\tilde{\theta}_{k}\|^2
\leq
\frac{\tilde{\theta}_{k}^T \bar{P}_{k}^{-1}\tilde{\theta}_{k}}
{\lambda_{\min}(\bar{P}_{k}^{-1})}
\notag \\
\leq&
\frac{1}{\lambda_{\min}(\bar{P}_{k}^{-1})} [
\tilde{\theta}_{k}\bar{P}_{k}^{-1}\tilde{\theta}_{k}
\!+\!(1\!-\!\frac{1}{\gamma}\!+\!o(1))\sum_{t=0}^{k-1}
\bar{a}_t(\tilde{\theta}_t^T\bar{\varphi}_t)^2 ]
\notag \\
=&
\frac{1}{\lambda_{\min}(\bar{P}_{k}^{-1})}[
O(1) + (1+\gamma+o(1))s_{k-1} \notag \\
&+ O(\ln r_{k-1} (\ln(e+ \ln r_{k-1}))^{\kappa 1_{\{\beta=2\}}}) ]
\notag \\
=&
O(\frac{1}{\lambda_{\min}(\bar{P}_{k}^{-1})})
+ (1+\gamma+o(1))\frac{s_{k-1}}{\lambda_{\min}(\bar{P}_{k}^{-1})}
\notag \\
&+ O(\frac{\ln r_{k-1} (\ln(e+ \ln r_{k-1}))^{\kappa 1_{\{\beta=2\}}}}
{\lambda_{\min}(\bar{P}_{k}^{-1})})
\quad \text{a.s. }.
\label{eq:DP_LS_thm_4_proof_1}
\end{align}
Note that $\bar{P}_{k}^{-1}$ is positive definite by \eqref{eq:DP_LS_alg_a}. Then, it follows from \eqref{ass:DP_LS_thm_4_pe_2} that
\begin{align}
0 < \frac{1}{\lambda_{\min}(\bar{P}_{k}^{-1})} \leq \frac{1}{\gamma_1k}
\quad \text{a.s. } ,
\label{eq:DP_LS_thm_4_proof_2}
\end{align}
which implies
\begin{align}
\frac{1}{\lambda_{\min}(\bar{P}_{k}^{-1})}\to 0,
\quad \text{a.s. } \quad \text{as} \quad k\to\infty.
\label{eq:DP_LS_thm_4_proof_3}
\end{align}
Note that
\begin{align*}
s_{k-1}
&=
\sum_{t=0}^{k-1}(\theta^T(\varphi_t-\bar{\varphi}_t))^2
\notag \\
&=
\sum_{t=0}^{k-1}(\sum_{j=1}^p a_j\eta_{t-j+1}
+ \sum_{i=1}^m\sum_{j=1}^{q_i}b_{i,j}\xi_{i,t-j+1})^2.
\end{align*}
Then, by Cauchy inequality and Assumption~\ref{ass:DP_LS_A2},
\begin{align}
s_{k-1}
&\leq
\|\theta\|^2
\sum_{t=0}^{k-1} (\sum_{j=1}^p \eta_{t-j+1}^2
+\sum_{i=1}^m\sum_{j=1}^{q_i}\xi_{i,t-j+1}^2)
\notag \\
&\leq
\|\theta\|^2
\sum_{t=0}^{k-1} (p\eta_{t}^2+\sum_{i=1}^m q_i \xi_{i,t}^2),
\label{eq:DP_LS_thm_4_proof_4}
\end{align}
where $\eta_{k+1} = 0, \xi_{i,k} =0$ for $k < 0$.

Consequently, by \eqref{eq:DP_LS_thm_4_proof_2},
\eqref{eq:DP_LS_thm_4_proof_4} and Kolmogorov's strong law of large numbers \cite{chow1988probability},
\begin{align}
&\lim_{k\to\infty} \frac{s_{k-1}}{\lambda_{\min}(\bar{P}_{k}^{-1})}
\leq
\lim_{k\to\infty} \frac{s_{k-1}}{\gamma_1 k}
\notag \\
\leq&
\lim_{k\to\infty} \frac{\|\theta\|^2
\sum_{t=0}^{k-1} (p\eta_{t}^2+\sum_{i=1}^m q_i \xi_{i,t}^2)}
{\gamma_1 k}
\notag \\
=&
\frac{\|\theta\|^2(2pb_0^2+2\sum_{i=1}^{m}q_ib_i^2)}{\gamma_1}
\quad \text{a.s. }.
\label{eq:DP_LS_thm_4_proof_5}
\end{align}
Therefore, letting $k\to\infty$ in \eqref{eq:DP_LS_thm_4_proof_1}, and
by \eqref{ass:DP_LS_thm_4_pe_1}, \eqref{eq:DP_LS_thm_4_proof_3}
and \eqref{eq:DP_LS_thm_4_proof_5}, we have
\begin{align*}
\lim_{k\to\infty} \|\tilde{\theta}_{k}\|^2
\leq
\|\theta\|^2
\frac{(1+\gamma)(2pb_0^2+2\sum_{i=1}^{m}q_ib_i^2)}{\gamma_1}
\quad \text{a.s. }.
\end{align*}
Let $\gamma\to 1^{+}$ in the above inequality. Then, the result is obtained.
\hfill $\blacksquare$

\begin{remark}
\eqref{ass:DP_LS_thm_4_pe_1}  in Theorem~\ref{thm:DP_LS_thm_4} are similar to the persistent excitation condition in {\rm \cite{chen1989convergence,chen1990identification,chen2014recursive}}.
In the case of $\beta>2$, \eqref{ass:DP_LS_thm_4_pe_1} reduces to
$\frac{\ln r_{k-1}}{\lambda_{\min}(\bar{P}_{k}^{-1})} \to 0$ which is equivalent
to the well-known possible weakest excitation condition for strong
consistency of the least-squares estimate for stochastic regressor model
{\rm \cite{lai1982least}}. It is worth pointing out that no independence and boundedness conditions on regression vectors are needed in Theorem ~\ref{thm:DP_LS_thm_4}.
\end{remark}

\begin{remark}
From the definition of $\bar{P}_{k}^{-1}$, $\gamma_1$ can be treated as a lower bound of the energy of regressor vector $\bar{\varphi}$. Besides, $s_{k-1}$ can be regarded as the energy of the added noises ($\eta_k$ and $\xi_{i,k}$). Then, $\frac{\gamma_1}{2pb_0^2+2\sum_{i=1}^m q_i b_i^2}$ can be regarded as the signal-to-noise ratio in some sense. \eqref{eq:DP_LS_thm_4_error_bound} shows that the estimation error of the algorithm is inverse to this signal-to-noise ratio.
\end{remark}

\begin{remark}
Theorem \ref{thm:DP_LS_thm_4} shows that the estimation error of the algorithm vanishes as $b_i$ vanishes for $i=0,1,\cdots,m$. However, \eqref{eq:DP_LS_thm_1_b0} and \eqref{eq:DP_LS_thm_1_b_i2} require $b_i$ to be larger if participants want to seek for better protection of the sensitive information, that is, smaller $\varepsilon$ and larger $\delta$.
\end{remark}
\begin{remark}
As shown in Theorem \ref{thm:DP_LS_thm_4}, in order to achieve the differential privacy, the algorithm does incur a loss on the utility, and thus gives biased estimates. This reveals a trade-off between the utility and the privacy of the algorithm, which is consistent with the current literature’s results on differentially private algorithms. The ability to achieve the differential privacy makes our approach suitable for privacy-critical scenarios.
\end{remark}

An interesting fact is that $\varepsilon$-differential privacy and almost sure convergence of the algorithm can be both achieved when there is no regression term in the output of System \eqref{eq:DP_LS_system} and only Participant $\mathcal{P}_0$ wants to protect its sensitive information.

\begin{theorem}
\label{thm:DP_LS_thm_5}
For System \eqref{eq:DP_LS_system} and Algorithm
\eqref{eq:DP_LS_alg_theta}-\eqref{eq:DP_LS_alg_P}, suppose that
$p=0$, $\xi_{i,k} =0$ for $i=1,\cdots,m$, $k\geq 0$, Assumptions~\ref{ass:DP_LS_A1},
\ref{ass:DP_LS_A3}, \ref{ass:DP_LS_A4}  and \eqref{eq:DP_LS_thm_1_b0}, \eqref{ass:DP_LS_thm_4_pe_1} hold. If there exists $\kappa>1$,
\begin{align}
\lambda_{\min}(\bar{P}_k^{-1}) \to \infty, \quad {\rm a.s.},
\label{eq:DP_LS_thm_5_1}
\end{align}
then
\renewcommand{\labelenumi}{\roman{enumi}.}
\begin{enumerate}
\item For Participant $\mathcal{P}_0$, Algorithm
\eqref{eq:DP_LS_alg_theta}-\eqref{eq:DP_LS_alg_P} is
$\varepsilon$-differentially private under $\delta$-adjacency.
\item $\theta_k \to \theta,$ {\rm a.s.}
\end{enumerate}
\end{theorem}
\noindent
{\bf Proof}. The first result directly follows from Theorem~\ref{thm:DP_LS_thm_1}. And so, it suffices to prove the second result.

Note that $p=0$ and $\xi_{i,k} =0$ for $i=1,\cdots,m$, $k\geq 0$,
$\varphi_k = \bar{\varphi}_k$ for $k\geq 0$. Then,
$d_k=\theta^T(\varphi_k-\bar{\varphi}_k)=0$ for $k\geq 0$,
and further, $s_k=0$ for $k\geq 0$.

By Lemma \ref{lem:DP_LS_lem_4} we have
\begin{align*}
&\quad \,
\tilde{\theta}_{k+1} \bar{P}_{k+1}^{-1} \tilde{\theta}_{k+1}
+ (1-\frac{1}{\gamma_1}+o(1))\sum_{t=0}^k
\bar{a}_t(\tilde{\theta}_t^T\bar{\varphi}_t)^2
\notag \\
&=
O(\ln r_k (\ln(e+ \ln r_k))^{\kappa 1_{\{\beta=2\}}}) + O(1)
\quad \text{a.s.}
\end{align*}
Therefore, for sufficiently large $k$,
\begin{align}
&\|\tilde{\theta}_{k}\|^2
\leq
\frac{\tilde{\theta}_{k}^T \bar{P}_{k}^{-1}\tilde{\theta}_{k}}
{\lambda_{\min}(\bar{P}_{k}^{-1})}
\notag \\
\leq&
\frac{1}{\lambda_{\min}(\bar{P}_{k}^{-1})} [
\tilde{\theta}_{k}\bar{P}_{k}^{-1}\tilde{\theta}_{k}
+(1-\frac{1}{\gamma_1}+o(1))\sum_{t=0}^{k-1}
\bar{a}_t(\tilde{\theta}_t^T\bar{\varphi}_t)^2 ]
\notag \\
=&
\frac{1}{\lambda_{\min}(\bar{P}_{k}^{-1})}[
O(1)
+ O(\ln r_{k-1} (\ln(e+ \ln r_{k-1}))^{\kappa 1_{\{\beta=2\}}}) ]
\notag \\
=&
O\!(\frac{1}{\lambda_{\min}(\bar{P}_{k}^{-1})}\!)
\!+\! O\!(\frac{\ln\!r_{k\!-\!1} (\ln(e\!+\! \ln r_{k\!-\!1}))^{\kappa 1_{\{\beta=2\}}}}
{\lambda_{\min}(\bar{P}_{k}^{-1})}\!)
\quad \text{a.s. }
\label{eq:DP_LS_thm_proof_1}
\end{align}
This together with  \eqref{ass:DP_LS_thm_4_pe_1} and \eqref{eq:DP_LS_thm_5_1} implies the second result.  \hfill $\blacksquare$

\begin{corollary}
For System \eqref{eq:DP_LS_system} and Algorithm
\eqref{eq:DP_LS_alg_theta}-\eqref{eq:DP_LS_alg_P}, suppose that
$p=0$, $\xi_{i,k} =0$ for $i=1,\cdots,m$, $k\geq 0$, Assumptions~\ref{ass:DP_LS_A1}, \ref{ass:DP_LS_A3}, \ref{ass:DP_LS_A4} and \eqref{eq:DP_LS_thm_1_b0} hold. If  there exist $\kappa>1$, $\gamma_2>0$ such that
\begin{align}
0< \limsup_{k > 0} \lambda_{\min}(\frac{1}{k}\bar{P}_k^{-1}) <\limsup_{k > 0} \lambda_{\max}(\frac{1}{k}\bar{P}_k^{-1}) < \gamma_2,
\quad {\rm a.s.},
\label{eq:DP_LS_thm_5_cor_1_1}
\end{align}
then
\renewcommand{\labelenumi}{\roman{enumi}.}
\begin{enumerate}
\item For Participant $\mathcal{P}_0$, Algorithm
\eqref{eq:DP_LS_alg_theta}-\eqref{eq:DP_LS_alg_P} is
$\varepsilon$-differentially private under $\delta$-adjacency.
\item $\theta_k \to \theta, $ {\rm a.s.} and
\begin{align*}
\|\tilde{\theta}_k\|^{2} = \left\{
\begin{array}{lc}
O(\frac{\ln k(\ln(e+\ln k))^{\kappa}}
{k}), & \beta=2; \\
O(\frac{\ln k}
{k}), &  \beta>2, 
\end{array} \right.
\quad {\rm a.s. }
\end{align*}
\end{enumerate}
\end{corollary}
\noindent
{\bf Proof}.
Note that \eqref{eq:DP_LS_thm_5_1} follows from
\eqref{eq:DP_LS_thm_5_cor_1_1} and $\ln r_k = O(\ln k)$ follows
from $r_k \leq 1+ tr{\bar{P}_{k+1}^{-1}} < 1
+ (p+\sum_{i=1}^m q_i)\lambda_{\max}(\bar{P}_{k+1}^{-1})$ and
\eqref{eq:DP_LS_thm_5_cor_1_1}. Then, the result yields from
\eqref{eq:DP_LS_thm_proof_1}. \hfill $\blacksquare$

\begin{remark} 
From Corollary 2, the proposed algorithm can only lead to consistent estimates if the system is exactly of the ARX form with the perturbations in the system output. However, as shown in Theorem \ref{thm:DP_LS_thm_4}, when the introduced perturbations are in both system output and system input,  the estimate of the algorithm does not converge to the true value $\theta$.  It should be pointed out that when there is no need for online computing, the total least square algorithm in the errors-in-variables problem  \cite{Torsten2007} can be used to overcome this problem. The estimate of the total least square algorithm can converge to the true value.
\end{remark}

\subsection{Optimize performance with guaranteed differential privacy}
For the convenience of the analysis, let
\begin{align*}
    \phi_k &= [y_k,\dots, y_{k-p+1}, u_{1,k},\dots, u_{1,k-q_1+1},\dots, \\
    &\quad\, u_{m,k},\dots, u_{m,k-q_m+1}]^T\in \mathbb{R}^{p+\sum_{i=1}^m q_i}, \\
    \zeta_k & = [\eta_k,\dots, \eta_{k-p+1}, \xi_{1,k},\dots, \xi_{1,k-q_1+1},\dots,\\
    &\quad\, \xi_{m,k},\dots, \xi_{m,k-q_m+1}]^T \in \mathbb{R}^{p+\sum_{i=1}^m q_i}.
\end{align*}
Then,
\begin{align}
   & \bar{P}_k^{-1} 
    = \sum_{t=0}^{k-1}(\bar{\phi}_t\bar{\phi}_t^T + \bar{P}_0^{-1}) \notag\\
    = &\sum_{t=0}^{k-1} \left[(\phi_t+\zeta_t)(\phi_t^T+\zeta_t)^T + \bar{P}_0^{-1} \right] \notag\\
    = &\sum_{t=0}^{k-1} \phi_t\phi_t^T \!+\! \sum_{t=0}^{k-1}\zeta_t\phi_t^T + \sum_{t=0}^{k-1}\phi_t\zeta_t^T \!+\! \sum_{t=0}^{k-1}\zeta_t\zeta_t^T \!+\! \bar{P}_0^{-1}.
    \label{eq:DP_LS_thm_4_proof_Pbar}
\end{align}
Based on Theorem \ref{thm:DP_LS_thm_4}, we have the following corollary.
\begin{corollary}
\label{thm:DP_LS_thm_6}
For System \eqref{eq:DP_LS_system} and Algorithm
\eqref{eq:DP_LS_alg_theta}-\eqref{eq:DP_LS_alg_P}, suppose that
Assumptions~\ref{ass:DP_LS_A1}-\ref{ass:DP_LS_A4},
\eqref{eq:DP_LS_thm_1_b0}, \eqref{eq:DP_LS_thm_1_b_i2}, \eqref{ass:DP_LS_thm_4_pe_1} hold. If 
\begin{align}
\sum_{t=1}^{k}\| \phi_t\|^2=O(k), {\rm a.s.},
\label{eq:DP_LS_thm_4as}
\end{align}
and there exists $\gamma_3>0$ such that $0<\gamma_3 < \liminf_{k\to\infty} \lambda_{\min}(P_k^{-1})/k$ with $P_k^{-1} = \sum_{t=0}^{k-1} \phi_t\phi_t^T + \bar{P}_0^{-1}$ being positive definite,
then
\renewcommand{\labelenumi}{\roman{enumi}.}
\begin{enumerate}
\item Algorithm
\eqref{eq:DP_LS_alg_theta}-\eqref{eq:DP_LS_alg_P} is $\varepsilon$-differentially private under $\delta$-adjacency for each participant;
\item For $\tilde{\theta}_k = \theta-\theta_k$, we have
\begin{align}
\limsup_{k\to\infty} \|\tilde{\theta}_k\|
\leq
2 \|\theta\| \sqrt{\frac{pb_0^2\!+\!\sum_{i=1}^{m}q_ib_i^2}
{\gamma_3 + 2\min_{i=0,1,\dots,m}{\{b_i^2\}}}},
\quad \!\! {\rm a.s.}
\label{eq:DP_LS_thm_4_error_bound1}
\end{align}
\end{enumerate}
\end{corollary}

\noindent
{\bf Proof}.
The $\varepsilon$-differential privacy follows from Theorem~\ref{thm:DP_LS_thm_4}. We only need to prove \eqref{eq:DP_LS_thm_4_error_bound1}.

From \eqref{eq:DP_LS_thm_4_proof_Pbar}, \eqref{eq:DP_LS_thm_4as} and Theorem 1.2.14 in  \cite{chen2014recursive} it follows that
\begin{align}
\bar{P}_k^{-1} = (1+o(1))P_k^{-1} + \sum_{t=0}^{k-1}\zeta_t\zeta_t^T.
   \label{eq:DP_LS_thm_4_proof_Pbar1}
\end{align}
By Kolmogorov's strong law of large numbers \cite{chow1988probability} and the property of Laplacian distribution, we have
\begin{align*}
    \frac{1}{k} \sum_{t=0}^{k-1} \zeta_t\zeta_t^T \to 2\mathrm{diag}(\underbrace{b_0^2,\dots,b_0^2}_{p},\underbrace{b_1^2,\dots, b_1^2}_{q_1},\dots,\underbrace{b_m^2,\dots, b_m^2}_{q_m}). \, \mathrm{a.s.}
\end{align*}
Note that $b_i > 0, i =0,1,\dots,m$, and  \eqref{eq:DP_LS_thm_4_proof_Pbar1}  holds. Then, 
$1/\lambda_{\min}(\bar{P}_k^{-1}) = O(\frac{1}{k}), \mathrm{a.s} .$.
Specifically, for any $\Delta \in(0, \min_{i=0,1,\dots, m}{\{b_i^2\}})$, there exists $k_0>0$ such that 
\begin{align}
    \frac{1}{\lambda_{\min}(\bar{P}_k^{-1})} \leq \frac{1}{(\gamma_3 + 2\min_{i=0,1,\dots, m}{\{b_i^2\}}-\Delta)k}, ~ \forall k>k_0,
    \label{eq:DP_LS_thm_4_proof_21}
\end{align}
which implies
\begin{align}
\frac{1}{\lambda_{\min}(\bar{P}_{k}^{-1})}\to 0,
\quad \text{a.s. } \quad \text{as} \quad k\to\infty.
\label{eq:DP_LS_thm_4_proof_31}
\end{align}
By \eqref{eq:DP_LS_thm_4_proof_4}, \eqref{eq:DP_LS_thm_4_proof_21},
 and Kolmogorov's strong law of large numbers,  we have 
\begin{align}
 &\limsup_{k\to\infty} \frac{s_{k-1}}{\lambda_{\min}(\bar{P}_{k}^{-1})}\notag \\
\leq&
\limsup_{k\to\infty} \frac{\|\theta\|^2
\sum_{t=0}^{k-1} (p\eta_{t}^2+\sum_{i=1}^m q_i \xi_{i,t}^2)}
{(\gamma_3 + 2\min_{i=0,1,\dots, m}{\{b_i^2\}}-\Delta)k}
\notag \\
=&\frac{\|\theta\|^2(2pb_0^2+2\sum_{i=1}^{m}q_ib_i^2)}{\gamma_3 + 2\min_{i=0,1,\dots, m}{\{b_i^2\}}-\Delta}
\quad \text{a.s. }
\label{eq:DP_LS_thm_4_proof_51}
\end{align}
Letting $k\to\infty$ in \eqref{eq:DP_LS_thm_4_proof_1},  by \eqref{ass:DP_LS_thm_4_pe_1}, \eqref{eq:DP_LS_thm_4_proof_31}
and \eqref{eq:DP_LS_thm_4_proof_51}, we have
\begin{align*}
\limsup_{k\to\infty} \|\tilde{\theta}_{k}\|^2
\leq
\|\theta\|^2
\frac{(1+\gamma)(2pb_0^2+2\sum_{i=1}^{m}q_ib_i^2)}{\gamma_3 + 2\min_{i=0,1,\dots, m}{\{b_i^2\}}-\Delta}
\quad \text{a.s. .}
\end{align*}
Let $\Delta\to 0^+, \gamma\to 1^+$ in the above inequality. Then, the result is obtained since $\Delta$ is arbitrary in $(0, \min_{i=0,1,\dots,m}{\{b_i^2\}})$ and $\gamma$ is arbitrary in $(1,\infty)$.
\hfill $\blacksquare$

From Corollary \ref{thm:DP_LS_thm_6}, the upper limit of the estimation error is bounded by the variances of added noises in the form of $2 \|\theta\| \sqrt{(pb_0^2\!+\!\sum_{i=1}^{m}q_ib_i^2)
/(\gamma_3 + 2\min_{i=0,1,\dots, m}{\{b_i^2\}})}$.
If participants choose $b_0$ and $b_i$ properly, then the upper limit of the estimation error can decrease. Next, we derive the following optimization problem for all participants to achieve $\varepsilon$-differential privacy and to minimize the bound of the estimation error's upper limit. 
\begin{align}
\!\!\!\!\!\underset{b_i:i=0,1,\cdots, m}{\arg\min}&
f(b_0,\cdots,b_m) = \frac{pb_0^2\!+\!\sum_{i=1}^{m}q_ib_i^2}
{\gamma_3 + 2\min_{i=0,1,\dots, m}{\{b_i^2\}}}
\label{eq:DP_LS_op_problem}
\\
\text{s.t.} \quad
&
(\ref{eq:DP_LS_thm_1_b0}), (\ref{eq:DP_LS_thm_1_b_i2}) \cr
&
b_i > 0, i=0,1,\cdots, m.\notag
\end{align}
\begin{lemma}
\label{lem:DP_LS_lem_5}
For System \eqref{eq:DP_LS_system} and Algorithm
\eqref{eq:DP_LS_alg_theta}-\eqref{eq:DP_LS_alg_P}, let $\varepsilon$
and $\delta$ be two positive numbers. Then, the optimum of the optimization problem \eqref{eq:DP_LS_op_problem} exists in a bounded set.
\end{lemma}
\noindent
{\bf Proof}. First, we show that the feasible set of the optimization problem is closed. Denote $v=(b_0,b_1,\dots,b_m) \in \mathbb{R}^{m+1}$ and  
$S_0 \equiv \{v\in \mathbb{R}^{m+1}|\frac{C_1\delta}{b_0}\leq \varepsilon, b_0 > 0\} = \{v\in \mathbb{R}^{m+1}| b_0\geq \frac{C_1\delta}{\varepsilon}>0 \}$, $S_i\equiv \{v\in \mathbb{R}^{m+1}|\frac{C_{i,2}}{b_0} + \frac{1}{b_i} \leq \frac{\varepsilon}{\delta}, b_0 > 0, b_i > 0\} \subset 
\{v\in \mathbb{R}^{m+1}|\frac{1}{b_i} \leq \frac{\varepsilon}{\delta}, \frac{C_{i,2}}{b_0} \leq \frac{\varepsilon}{\delta}, b_i > 0, b_0> 0\}
=\{v\in \mathbb{R}^{m+1}| b_0\geq \frac{\delta}{\varepsilon} > 0, b_i \geq \frac{C_{i,2}\delta}{\varepsilon} > 0\}\equiv S_i^{'}$. Note that $S'_i$ is a closed set and $g(b_0, b_i) \equiv \frac{C_{i,2}}{b_0}+\frac{1}{b_i}$ is continuous on $S'_i$. Then, $g(b_0,b_i)$'s preimage of $[\varepsilon, \infty)$, i.e. $S_i$, is a closed set.

By the definition of $S_i$, the feasible set of the optimization problem \eqref{eq:DP_LS_op_problem} is $\bigcap_{i=0}^m S_i$, which is a closed set. 

Next, we show that there exists a positive number $r^* > 0$ and a bounded cube $N(r^*) \equiv \{v\in \mathbb{R}^{m+1}|0 < b_i \leq r^*,i=0,1,\dots,m \}$ such that the optimal solution $v^* \equiv (b_0^*, b_1^*,\dots, b_m^*) \in (\bigcap_{i=0}^m S_i) \bigcap N(r^*)$. Below the analysis is undertaken by two cases.

{\bf Case 1:  there exists $i_0\!\in\!\{1,\!2,\!\cdots\!,\!m\}$ satisfying $C_{i_0,2} \geq C_1$.}

Since \eqref{eq:DP_LS_thm_1_b0} is implied by \eqref{eq:DP_LS_thm_1_b_i2} indexed by $i_0$,  \eqref{eq:DP_LS_thm_1_b0} can be omitted, i.e. $\bigcap_{i=0}^m S_i = \bigcap_{i=1}^m S_i$. 

Rewrite \eqref{eq:DP_LS_thm_1_b_i2} as $\frac{C_{i,2}}{b_0}+ \frac{1}{b_{i}}\leq \frac{\varepsilon}{\delta}.$ Then, for any $v=(b_0,b_1,\dots,b_m)\in \bigcap_{i=1}^m S_i$, there always exists $\Delta_{i,1} \in (0,1)$ such that 
\begin{align}
\frac{C_{i,2}}{b_0} = (1-\Delta_{i,1})\frac{\varepsilon}{\delta}\quad\text{and}\quad \frac{1}{b_{i}} \leq \frac{\Delta_{i,1}\varepsilon}{\delta}.
\label{eq:DP_LS_op_proof_1}
\end{align}

Let $i_0 = \arg\max_i\{C_{i,2}\}$. Then, $\Delta_{i,1}\geq \Delta_{i_0,1}$ for $i\in\{1,2,\dots,m\} \setminus \{i_0\}$.  By choosing $\Delta_{i_0,0} =\min\{\frac{1}{1+C_{i_0,2}},\sqrt{\frac{q_{i_0}}{4pC_{i_0,2}}}, \frac{1}{2}\}$, 
we can claim that $v$ is not optimal for \eqref{eq:DP_LS_op_problem}, 
if $\Delta_{i_0,1}$ in \eqref{eq:DP_LS_op_proof_1} is strictly smaller than $\Delta_{i_0,0}$.
By this claim, for the optimal solution $v^*$, we have $\frac{1}{b_i^*} \geq \frac{\Delta_{i_0,0}\varepsilon}{\delta}$, i.e. $b_i^*\leq \frac{\delta}{\Delta_{i_0,0}\varepsilon}$ for $i=1,2,\dots,m$.

Below we will explain how this claim is established. Let 
$ \frac{C_{i_0,2}}{b_0'} = (1-\Delta_{i_0,0})\frac{\varepsilon}{\delta}$ and $\frac{1}{b_{i_0}'}= \frac{\Delta_{i_0,0},\varepsilon}{\delta}. $
Then, from $\Delta_{i_0,1} < \Delta_{i_0,0}$  it follows that 
\begin{align*}
b_{i_0}' = \delta/(\Delta_{i_0,0}\varepsilon) \leq b_{i_0}
\quad \text{and} \quad
b_0' = \frac{C_{i_0,2}\delta}{(1-\Delta_{i_0,0})\varepsilon} > b_0.
\end{align*}
Denote $v'=(b_0',\dots,b_{i_0-1},b_{i_0}',b_{i_0+1},\dots,b_m)$.
Then, by $\frac{C_{i_0,2}}{b_0'}\leq \frac{C_{i_0,2}}{b_0}$ we have $v'\in \bigcap_{i=1}^m S_i$.

Denote $\Delta_{i_0,0}-\Delta_{i_0,1}$ by $\alpha \Delta_{i_0,0}$ for some $\alpha \in [0,1)$. Then,
\begin{align}
    b_0-b_0'&=\frac{\alpha\Delta_{i_0,0}}{(1-\Delta_{i_0,0})[1-(\Delta_{i_0,0}-\alpha\Delta_{i_0,0})]}\frac{C_{i_0,2}\delta}{\varepsilon} 
    \notag \\
    &= \frac{\alpha \Delta_{i_0,0}}{1-(2-\alpha)\Delta_{i_0,0}+(1-\alpha)\Delta_{i_0,0}^2}\frac{C_{i_0,2}\delta}{\varepsilon}
    \label{eq:DP_LS_op_proof_2}\\
    b_{i_0}-b_{i_0}' &= \frac{1}{\Delta_{i_0,0 }}\frac{\alpha}{1-\alpha}\frac{\delta}{\varepsilon}.
    \label{eq:DP_LS_op_proof_3}
\end{align}
Denote $m_1 = \min\{ b_0^2,\dots,b_{i_0-1}^2,b_{i_0}^2,b_{i_0+1}^2,\dots,b_m^2 \}$ and  
$m_0 = \min\{b_0'^2,\dots, b_{i_0-1}^2,b_{i_0}'^2,b_{i_0+1}^2,\dots,b_m^2 \}$. 
Note that $\Delta_{i_0,1}\leq \Delta_{i_0,0}\leq \frac{1}{1+C_{i_0,2}}$ implies $b_0'< b_{i_0}'$. Then, together with $b_{i_0}'<b_{i_0}$ and $b_0' > b_0$, we have $m_1\leq m_0$, and  further 
\begin{align}
f(v)
&= \frac{pb_0^2+q_{i_0}b_{i_0}^2+\sum_{j\neq i_0,j\geq 1}^{m}q_jb_j^2}
{\gamma_1 + 2m_1}\notag\\
&\geq \frac{pb_0^2+q_{i_0}b_{i_0}^2+\sum_{j\neq i_0,j\geq 1}^{m}q_jb_j^2}
{\gamma_1 + 2m_0}\notag\\
&= f(v') + \frac{p(b_0^2-b_0'^2)+q_{i_0}(b_{i_0}^2-b_{i_0}'^2)}{\gamma_1+2m_0} \label{eq:DP_LS_op_proof_5}
\end{align}
By \eqref{eq:DP_LS_op_proof_2} and \eqref{eq:DP_LS_op_proof_3} we have 
\begin{align*}
    \frac{b_{i_0}-b_{i_0}'}{b_0'-b_0}
    &=\frac{1-(2-\alpha)\Delta_{i_0,0}+(1-\alpha)\Delta_{i_0,0}^2}{(1-\alpha)C_{i_0,2}\Delta_{i_0,0}^2}  
\end{align*}
Since $\alpha \in [0,1)$, the polynomial $1-(2-\alpha)\Delta_{i_0,0}+(1-\alpha)\Delta_{i_0,0}^2$ strictly decreases on the interval $(0,1)$ with respect to $\Delta_{i_0,0}$. Note that $\Delta_{i_0,0}\leq \frac{1}{2}$. Then, 
$1-(2-\alpha)\Delta_{i_0,0}+(1-\alpha)\Delta_{i_0,0}^2\geq \frac{1+\alpha}{4}\geq \frac{1}{4}$. Thus, by $\Delta_{i_0,0} \leq \sqrt{\frac{q_{i_0}}{4pC_{i_0,2}}}$ we obtain
\begin{align*} 
\frac{b_{i_0}-b_{i_0}'}{b_0'-b_0}
&\geq \frac{1}{4C_{i_0,2}\Delta_{i_0,0}^2}\geq \frac{p}{q_{i_0}}\geq \frac{p(b_0+b_0')}{q_{i_0}(b_{i_0}+b_{i_0}')},
\end{align*}
which means $p(b_0^2-b_0'^2)+q_{i_0}(b_{i_0}^2-b_{i_0}'^2)\geq 0$. 
This together with \eqref{eq:DP_LS_op_proof_5} implies that 
$f(v)\geq f(v')$. The former claim is established.

For any $v=(b_0,b_1,\dots,b_m)\in \bigcap_{i=1}^m S_i$, there always exists $\Delta_{i,2} \in (0,1)$ such that 
\begin{align}
    \frac{C_{i,2}}{b_0} \leq \Delta_{i,2}\frac{\varepsilon}{\delta}\quad\text{and}\quad \frac{1}{b_{i}}= (1-\Delta_{i,2})\frac{\varepsilon}{\delta}.
    \label{eq:DP_LS_op_proof_6}
\end{align}
For any $i=1,2,\dots,m$, by properly choosing a small positive number $\Delta_{i,0}'$, we can claim that the feasible solution $v$ is not optimal, if $\Delta_{i,2}$ in \eqref{eq:DP_LS_op_proof_6} is strictly less than $\Delta_{i,0}'$. By this claim, for the optimal solution $v^* $, we have $\frac{C_{i,2}}{b_0^*}\geq \Delta_{i,0}'\frac{\varepsilon}{\delta}$ for $i=1,2,\dots,m$, which implies $b_0^* \leq \frac{\delta}{\varepsilon}\max_{1\leq i\leq m}\{\frac{C_{i,2}}{\Delta_{i,0}'}\}$.

Let $b_0'(i) = \frac{C_{i,2}\delta}{\Delta_{i,2}\varepsilon}$. Then,
\begin{align*}
    |b_0-b_0'(i)| \leq \frac{|\Delta_{i,0}'-\Delta_{i,2}|}{\Delta_{i,0}'\Delta_{i,2}}\frac{C_{i,2}\delta}{\varepsilon}.
\end{align*}
Similar to the proof of the former claim, this claim holds.

{\bf Case 2:  $C_{i,2} < C_1$ for all $i=1,2\dots,m$.}

Note that $\frac{C_{i,2}\delta}{b_0} < \frac{C_1\delta}{b_0}\leq \varepsilon$. Then, $b_0\geq \frac{C_1\delta}{\varepsilon}$ by \eqref{eq:DP_LS_thm_1_b0}, and $b_i\geq \frac{C_1\delta}{(1-C_{i,2})\varepsilon}$ by \eqref{eq:DP_LS_thm_1_b_i2}.  When $b_0=b_1=\dots=b_m = \bar{b}$, $f(\bar{b},\dots,\bar{b}) = \frac{\bar{b}^2(p+\sum_{i=1}^m q_i)}{\gamma_1+\bar{b}^2}$ strictly increases in $\bar{b}$. Thus, for any feasible solution $v=(b_0,b_1,\dots,b_m)$ with $b_i \geq \max\{\frac{C_1\delta}{\varepsilon}, \max_{1\leq i\leq m}\{\frac{C_1\delta}{(1-C_{i,2})\varepsilon}\}\} \equiv \bar{b}_0$ for $i=1,2,\dots,m$, we can always decrease the elements of $v$ to its mimimum one, and then to $\bar{b}_0$. By \eqref{eq:DP_LS_op_problem} we have $f(v)\geq f(\bar{b}_0,\dots,\bar{b}_0)$, which implies $b^*$ is bounded in $N(\bar{b}_0)$. \hfill$\blacksquare$

Based on the above lemma and Theorems \ref{thm:DP_LS_thm_1}, \ref{thm:DP_LS_thm_2}, Corollary \ref{thm:DP_LS_thm_6}, we have the following theorem.
\begin{theorem}
\label{thm:DP_LS_thm_7}
For System \eqref{eq:DP_LS_system} and Algorithm \eqref{eq:DP_LS_alg_theta}-\eqref{eq:DP_LS_alg_P}, suppose that Assumptions~\ref{ass:DP_LS_A1}-\ref{ass:DP_LS_A4} and \eqref{ass:DP_LS_thm_4_pe_1}, \eqref{eq:DP_LS_thm_4as} hold. If there exists $\gamma_3>0$ such that  $0<\gamma_3 < \liminf_{k\to\infty} \lambda_{\min}(P_k^{-1})/k$ with $P_k^{-1} = \sum_{t=0}^{k-1} \phi_t\phi_t^T + \bar{P}_0^{-1}$ being positive definite,  and each participant chooses $b_i$ according to \eqref{eq:DP_LS_thm_1_b0} and \eqref{eq:DP_LS_thm_1_b_i2}
for $i=0,1,\cdots,m$, then Algorithm \eqref{eq:DP_LS_alg_theta}-\eqref{eq:DP_LS_alg_P} is $\varepsilon$-differentially private under $\delta$-adjacency for
each participant, and there exists $f^*$ such that $\lim_{k\to\infty} \|\tilde{\theta}_k\| \leq
2\|\theta\| \sqrt{f^*},$ where $f^*$ is the minimal value of the optimization problem \eqref{eq:DP_LS_op_problem}.
\end{theorem}

\begin{remark}
For given privacy indexes $\varepsilon$ and $\delta$, Theorem \ref{thm:DP_LS_thm_7}
shows the existence of  the noise parameters $b_i$'s to minimize
the estimation error of the algorithm with guaranteed differential privacy.
\end{remark}

\section{Numerical examples}
In this section, we give two examples to show the efficiency of the proposed algorithm. 

{\bf Example 1}. Consider the following MP-ARX systems with 4 participants:
\begin{align}
y_{k+1}
=& -\frac{1}{4} y_k + \frac{3}{8} y_{k-1}+ u_{1,k} + 2u_{1,k-1}\cr
&+ 3u_{2,k} + 4u_{2,k-1}+ 5u_{3,k} + 6u_{3,k-1}+ \omega_{k+1},
\label{eq:DP_LS_sim_sys}
\end{align}
where $\omega_k \sim N(0,1)$. Then, we have $p=q_1=q_2=q_3=2$,
$A = \begin{bmatrix}
    0& 1\\
    \frac{3}{8} & -\frac{1}{4} 
   \end{bmatrix}$, and $\theta = [-\frac{1}{4}, \frac{3}{8}, 1, 2, 3, 4, 5, 6]^T.$ The characteristic polynomial of System \eqref{eq:DP_LS_sim_sys} is
$\lambda(z) = 1 + \frac{1}{4} z - \frac{3}{8} z^2$ with characteristic
roots $z_1 = 2, z_2 = -\frac{4}{3}$ that lie outside the unit circle,
and thus, System \eqref{eq:DP_LS_sim_sys} is asymptotically stable. Note that by Jordan decomposition, $A = S J S^{-1}$ with
$J = \begin{bmatrix}
-\frac{3}{4} & 0 \\
0 & \frac{1}{2}
\end{bmatrix}$ and
$S = \begin{bmatrix}
\frac{3}{5} & \frac{1}{\sqrt{5}} \\
-\frac{4}{5} & \frac{2}{\sqrt{5}}
\end{bmatrix}$. Then, by Remark~\ref{remark lS 2},
$\|A^k\|\leq 1.7 \times (\frac{3}{4})^k$ with $c_0 = \|S\|\|S^{-1}\| \approx 1.618,
\lambda = \frac{3}{4}$. Hence, by
\eqref{eq:DP_LS_C1} we have $C_1 \approx 7.864$,
by \eqref{eq:DP_LS_C_i2} we have $C_{1,2} \approx 23.594$,
$C_{2,2} \approx 55.053$ and $C_{3,2} \approx 86.512$.
Next, set ${\bar{P}_0}^{-1} = I$, $\theta_0 = [0, 0, 0, 0, 0, 0, 0, 0]^T$.

\begin{figure}[httb]
\centering
\includegraphics[width=0.38\textwidth]{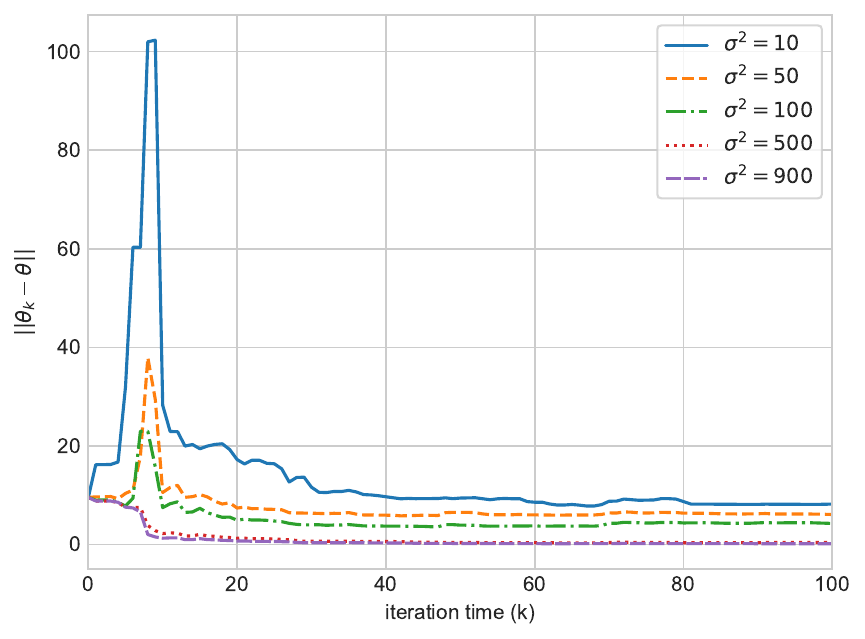}
\caption{Estimation error of the algorithm under different inputs}
\label{fig:DP_LS_input_1}
\end{figure}

To show the performance of the algorithm, set
$\varepsilon = 0.5, \delta =1$, $u_{i,k} \sim N(0, \sigma^2)$
with $\sigma^2 = 10, 50, 100, 500, 900$, respectively.
Then, the estimation error of the algorithm under different inputs is
shown in Fig.~\ref{fig:DP_LS_input_1}. From Fig.~
\ref{fig:DP_LS_input_1} it follows that the estimation error of the
algorithm exists due to the added privacy-preserving noises. But as
the input becomes more informative, i.e. $\sigma^2$ increases, this
error can be reduced to an acceptable level.

\begin{figure}[httb]
\centering
\includegraphics[width=0.38\textwidth]{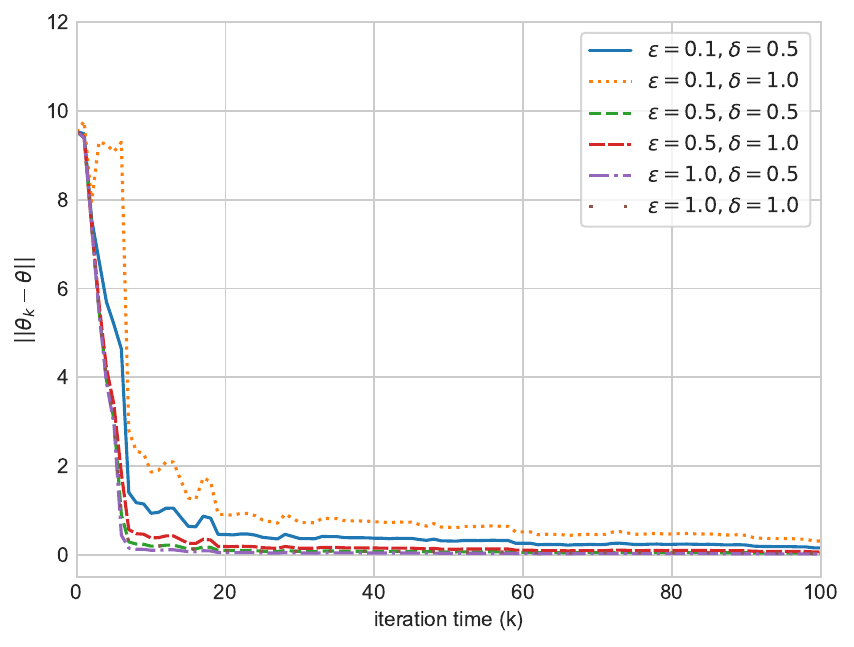}
\caption{Estimation error of the algorithm under different privacy indexes when $p=\xi_{1,k}=\xi_{2,k} = 0$}
\label{fig:DP_LS_dpls_vs_ols}
\end{figure}

To show the performance of the algorithm for the special case in
Theorem \ref{thm:DP_LS_thm_5}, we remove the regression terms ($y_k$ and
$y_{k-1}$) in \eqref{eq:DP_LS_sim_sys} and set
$\xi_{i,k}=0$, $u_{i,k}\sim N(0, 100)$, $i=1,2$, $\varepsilon = 0.1, 0.5, 1$, $\delta = 0.5, 1$, respectively. Then, the estimation error of the algorithm under
different privacy indexes is shown in Fig. \ref{fig:DP_LS_dpls_vs_ols}.
From Fig. \ref{fig:DP_LS_dpls_vs_ols} it follows that the estimation
error of the algorithm decreases to 0 no matter what values $\varepsilon$
and $\delta$ take on.

\begin{figure}[httb]
\centering
\subfigure[Different $\varepsilon$]{\includegraphics[width=0.24\textwidth]{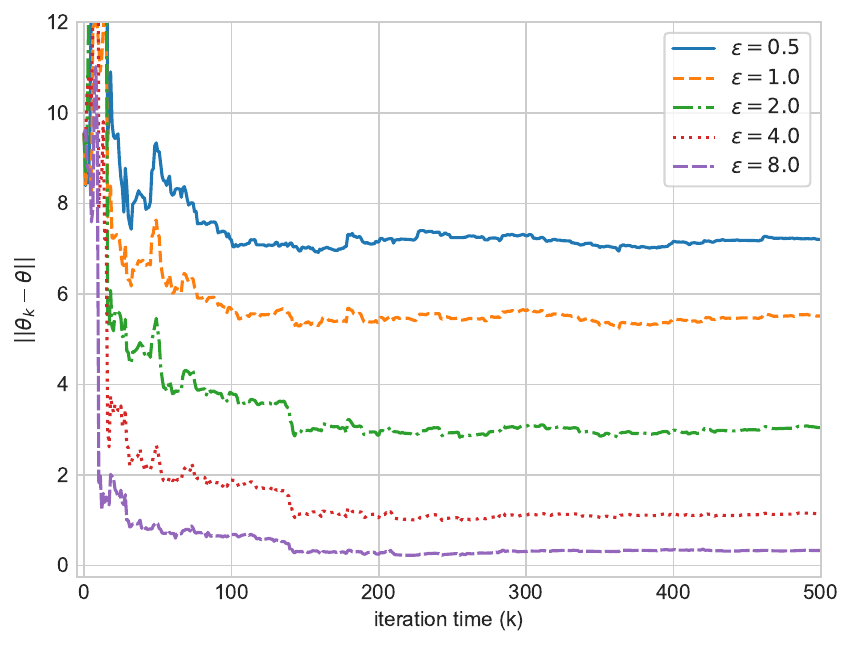}}
\subfigure[Different $\delta$]{\includegraphics[width=0.24\textwidth]{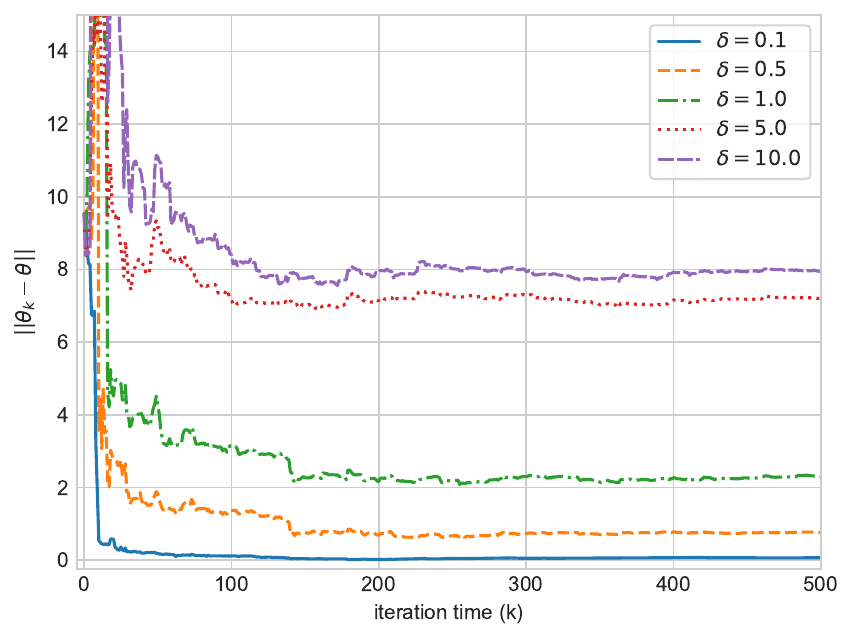}}
\caption{Estimation error of the algorithm under different variances of the added Laplacian noise}
\label{fig:DP_LS_epsilon}
\end{figure}

Next, we show the influence of $\varepsilon$ and $\delta$ on the
estimation error of the algorithm, respectively. Let $\delta = 1, u_{i,k} \sim N(0, 10)$
be fixed, and choose
$\varepsilon = 0.5, 1, 2, 4, 8$, respectively. Then, the estimation
error of the algorithm under different $\varepsilon$ is shown in
Fig. \ref{fig:DP_LS_epsilon} (a). From Fig.
\ref{fig:DP_LS_epsilon} (a) it follows that for given $\delta$, the
larger $\varepsilon$ is, the smaller the estimation error is.
Let $\varepsilon = 0.5, u_{i,k} \sim N(0, 100)$ be fixed,
and choose $\delta = 0.1, 0.5, 1, 5, 10$, respectively.
Then, the estimation error of the algorithm under different $\delta$
is shown in Fig. \ref{fig:DP_LS_epsilon} (b).
From Fig. \ref{fig:DP_LS_epsilon} (b) it follows that for given
$\varepsilon$, the smaller $\delta$ is, the smaller the estimation
error is.  Hence, from Fig. \ref{fig:DP_LS_epsilon} it follows that 
the smaller the noise variance (the larger $\varepsilon$ in Fig. \ref{fig:DP_LS_epsilon} (a) or the smaller $\delta$ in Fig. \ref{fig:DP_LS_epsilon} (b)) is, the smaller the estimation error is.

\begin{figure}[httb]
\centering
\includegraphics[width=0.38\textwidth]{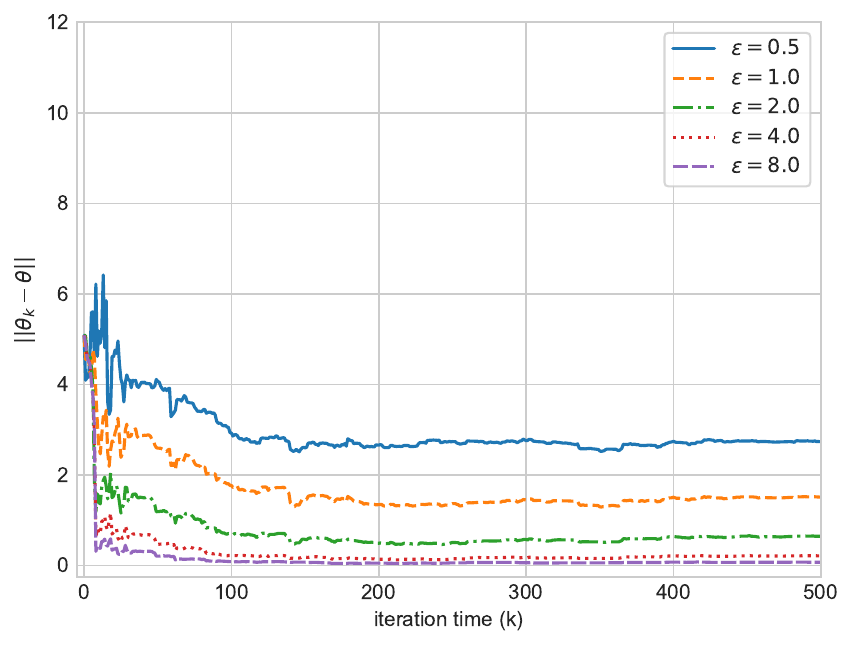}
\caption{Estimation error of the algorithm under different $\varepsilon$}
\label{fig:DP_LS_epsilon1}
\end{figure}

{\bf Example 2}.  In econometric research fields, we study the impact of historical investment behaviors of 3 banks on the economic development situation while protecting the sensitive information involved. $y_{k}$ represents the economic development situation of the $k$th month (internal evaluation indicators needed to be protected), and $u_{i,k}$ represents the credit investment scale of the $i$th bank in the $k$th month. $\omega_{k+1}$ represents the uncertainty in the $k+1$th month.  Both economic development situation and  credit investment scale of each bank are sensitive information. The following MP-ARX systems with 4 participants are given to investigate the relationship between the economic development situation of the $k+1$th month and the historical investment scale of each bank. 
\begin{align*}
y_{k+1}=& -\frac{1}{4} y_k + \frac{3}{8} y_{k-1}+ 2u_{1,k} +2.2u_{1,k-1}+ 1.5u_{2,k}\cr
&+ 2.5u_{2,k-1}+ 2.4u_{3,k} + 1.6u_{3,k-1}+ \omega_{k+1},
\end{align*}
We show  the estimation error of the algorithm under different $\varepsilon$. Let $\delta = 1, u_{i,k} \sim N(0, 10)$ be fixed, and choose
$\varepsilon = 0.5, 1, 2, 4, 8$, respectively. Then, the estimation
error of the algorithm under different $\varepsilon$ is shown in
Fig. \ref{fig:DP_LS_epsilon1}. From Fig.
\ref{fig:DP_LS_epsilon1} it follows that the
larger $\varepsilon$ is, the smaller the estimation error is. This is consistent with the theoretical analysis.
\section{Conclusions}
This paper has proposed a differentially private recursive least-squares algorithm for MP-ARX systems. A rigorous mathematical proof of differential privacy of the algorithm is established when the system is asymptotically stable, and well-designed noises are introduced to protect participants' sensitive information. We show that the asymptotic stability of the system is necessary for ensuring the differential privacy of the algorithm. The estimation error and convergence rate of the algorithm are provided under the general and possible weakest excitation condition without requiring the boundedness, independence and stationarity on the regression vectors. In particular, if there is no regression term in the system output and the differential privacy only on the system output is considered, then the $\varepsilon$-differential privacy and almost sure convergence of the algorithm can be achieved simultaneously. Furthermore,  we prove the  existence of the added noise intensity to minimize the estimation error of the algorithm with $\varepsilon$-differential privacy. 

The algorithm proposed in this paper is not unbiased. In the future work, it is worth considering how to design a recursive identification algorithm for MP-ARX systems that can simultaneously achieve unbiased estimates and differential privacy for each participant.

\begin{IEEEbiography}
[{\includegraphics[width=1in,height=1.25in,clip,keepaspectratio]{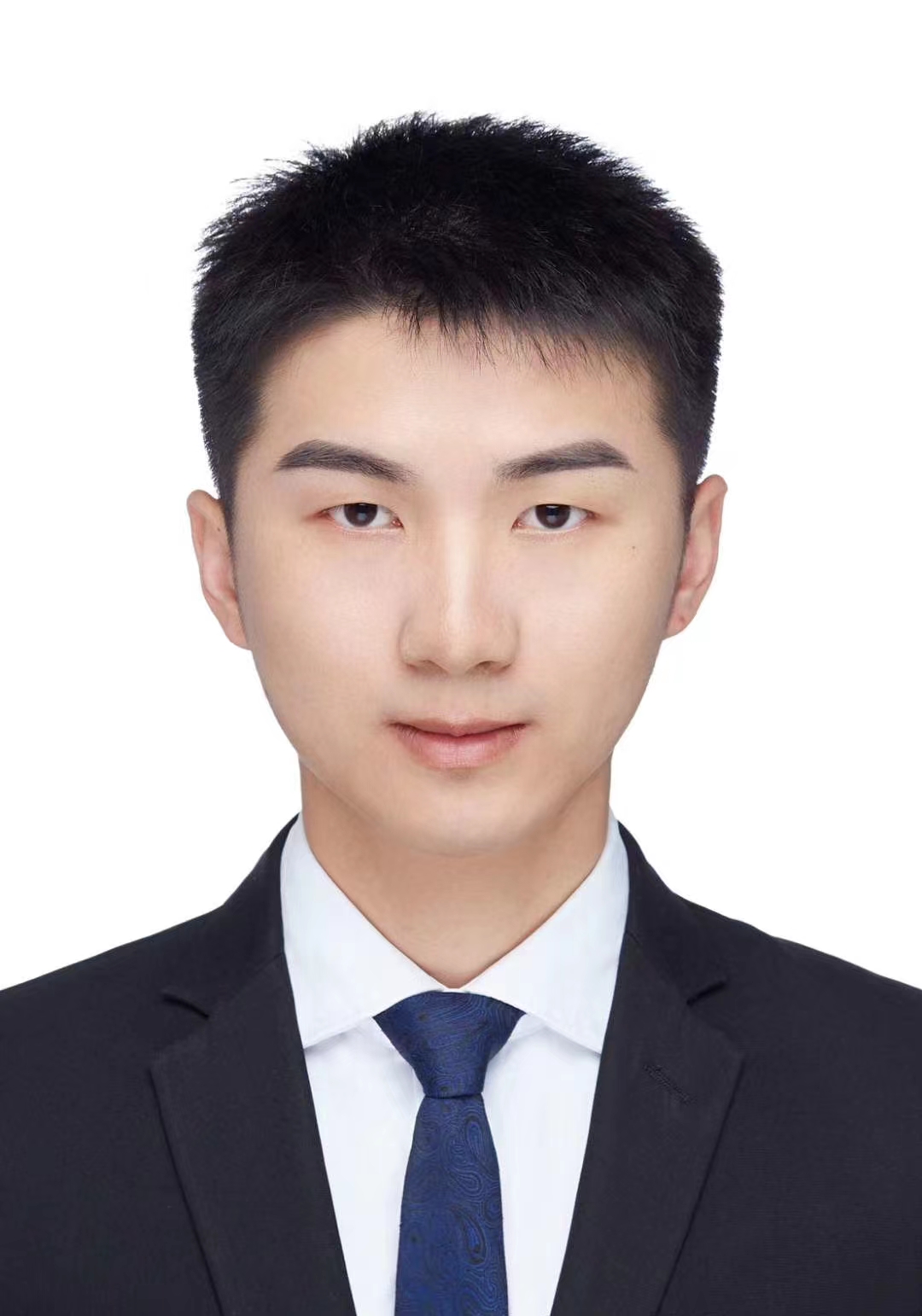}}]{Jianwei Tan}
received the B.S. degrees in Automation and Applied Mathematics from Nankai University, Tianjin, China, in 2015 and the Ph.D. degree in Systems Theory from Academy of Mathematics and and Systems Science, Chinese Academy of Science, Beijing, China, in 2023. He is currently an Engineer at Shanghai Aerospace Control Technology Institute. His current research interests include system identification, privacy preservation for uncertain systems.
\end{IEEEbiography}

\begin{IEEEbiography}
[{\includegraphics[width=1in,height=1.25in,clip,keepaspectratio]{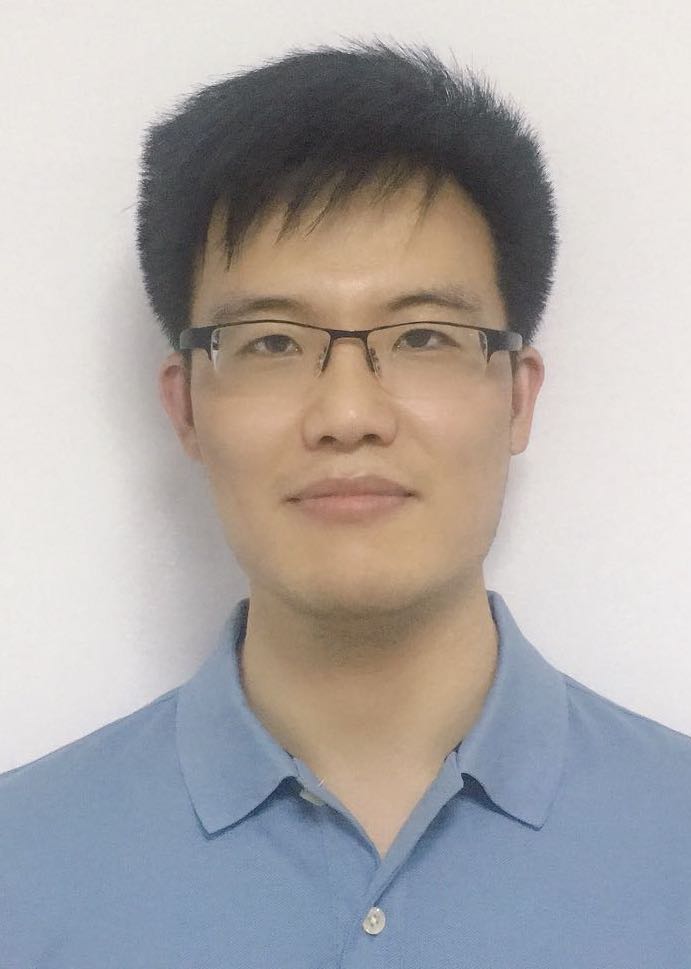}}]{Jimin Wang}
received the B.S. degree in mathematics from Shandong Normal University, China, in 2012 and the Ph.D. degree from the School of Mathematics, Shandong University, China, in 2018. From May 2017 to May 2018, he was a joint Ph.D. student with the School of Electrical Engineering and Computing, The University of Newcastle, Australia. From July 2018 to December 2020, he was a postdoctoral researcher in the Institute of Systems Science (ISS), Chinese Academy of Sciences (CAS), China. He is currently an associate professor in the School of Automation and Electrical Engineering, University of Science and Technology Beijing. His current research interests include privacy and security in cyber-physical systems, stochastic systems and networked control systems. 

He is a member of the IEEE CSS Technical Committee on Security and Privacy,  the IEEE CSS Technical Committee on Networks and Communication Systems, the IFAC Technical Committee 1.5 on Networked Systems. He was a recipient of Shandong University's excellent doctoral dissertation. 
\end{IEEEbiography}

\begin{IEEEbiography}
[{\includegraphics[width=1in,height=1.25in,clip,keepaspectratio]{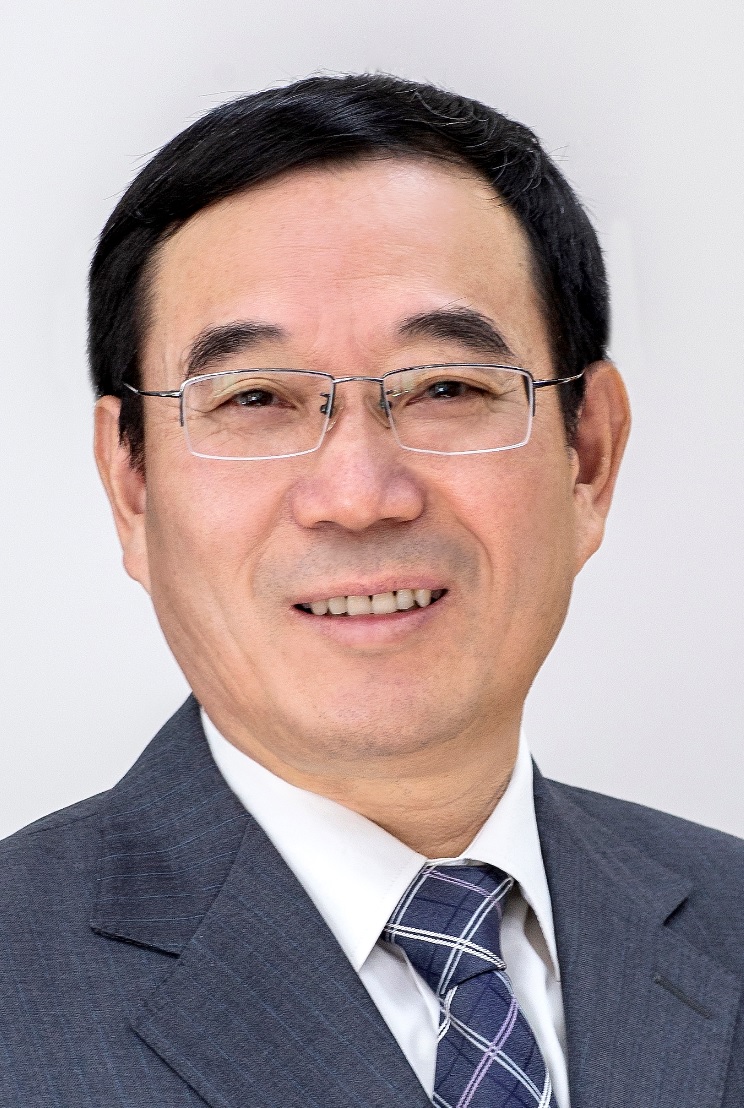}}]{Ji-Feng Zhang}
received the B.S. degree in mathematics from Shandong University, China, in 1985 and the Ph.D. degree from the Institute of Systems Science (ISS), Chinese Academy of Sciences (CAS), China, in 1991. Since 1985, he has been with the ISS, CAS. Now he is also with the School of Automation and Electrical Engineering, Zhongyuan University of Technology. His current research interests include system modeling, adaptive control, stochastic systems, and multi-agent systems.

He is an IEEE Fellow, IFAC Fellow, CAA Fellow, CSIAM Fellow, member of the European Academy of Sciences and Arts, and Academician of the International Academy for Systems and Cybernetic Sciences. He received the second prize of the State Natural Science Award of China in 2010 and 2015, respectively. He was a Vice-Chair of the IFAC Technical Board, member of the Board of Governors, IEEE Control Systems Society; Convenor of Systems Science Discipline, Academic Degree Committee of the State Council of China; Vice-President of the Systems Engineering Society of China, the Chinese Mathematical Society, and the Chinese Association of Automation. He has served as Editor-in-Chief, Deputy Editor-in-Chief, Senior Editor or Associate Editor for more than 10 journals, including {\it Science China Information Sciences},  {\it National Science Review},  {\it IEEE Transactions on Automatic Control}, and  {\it SIAM Journal on Control and Optimization} etc.
\end{IEEEbiography}

\end{document}